\documentclass[10pt]{article}
\pdfoutput=1
\usepackage{jheppub,amsmath,amssymb}
\usepackage{enumerate}
\usepackage{bm}
\usepackage{bbm}
\usepackage[usenames,dvipsnames]{xcolor}
\usepackage{graphics}
\usepackage{tikz,todonotes}
\usepackage[utf8]{inputenc}
\usepackage{cancel}
\usetikzlibrary{arrows,positioning,decorations.markings,decorations.pathmorphing,calc}
\usepackage{color}
\usepackage{graphicx}
\usepackage{dcolumn}
\usepackage{bm}
\usepackage{amssymb}
\usepackage{amsmath}
\usepackage{sectsty}
\usepackage{colortbl}
\usepackage{latexsym}
\usepackage{float}
\usepackage{ifthen}
\usepackage{caption,subfig}
\usepackage{enumerate}
\usepackage{url}
\usepackage{caption,subfig}
\usepackage{xfrac}
\usepackage{amsmath}

\DeclareFontFamily{U}{mathb}{\hyphenchar\font45}
\DeclareFontShape{U}{mathb}{m}{n}{
      <5> <6> <7> <8> <9> <10> gen * mathb
      <10.95> mathb10 <12> <14.4> <17.28> <20.74> <24.88> mathb12
      }{}
\DeclareSymbolFont{mathb}{U}{mathb}{m}{n}
\DeclareFontSubstitution{U}{mathb}{m}{n}

\let\dot\relax
\DeclareMathAccent{\dot}{0}{mathb}{"39}
\let\ddot\relax
\DeclareMathAccent{\ddot}{0}{mathb}{"3A}
\let\dddot\relax
\DeclareMathAccent{\dddot}{0}{mathb}{"3B}
\let\ddddot\relax
\DeclareMathAccent{\ddddot}{0}{mathb}{"3C}

\definecolor{hyperref}{RGB}{026,028,087}

\newcommand{\Lic}{{Lichnerowicz}}
\newcommand{\mpl}{M_{\rm Pl}}

\def\be{\begin{equation}}
\def\ee{\end{equation}}
\def\ba{\begin{eqnarray}}
\def\ea{\end{eqnarray}}

\def\nn{\nonumber}

\def\d{\mathrm{d}}

\def\ba{\begin{eqnarray}}
\def\ea{\end{eqnarray}}

\def\O{\mathcal{O}}

\def\L{\mathcal{L}}

\def\A{\mathcal{A}}

\def\d{\mathrm{d}}
\def\mn{_{\mu \nu}}
\def\ab{_{\alpha \beta}}

\def\({\left(}
\def\){\right)}
\def\e{\epsilon}

\def\mpl{M_{\rm Pl}}
\def\p{\partial}

\def\Ns{N^*_0}
\def\Nv{N^*_1}
\def\Np{N^*_{1/2}}

\usepackage{feyn}

\begin{document}

\title{The Speed of Gravity}

\author{Claudia de Rham}
\author{\& Andrew J. Tolley}
\affiliation{Theoretical Physics, Blackett Laboratory, Imperial College, London, SW7 2AZ, UK}
\affiliation{CERCA, Department of Physics, Case Western Reserve University, 10900 Euclid Ave, Cleveland, OH 44106, USA}

\emailAdd{c.de-rham@imperial.ac.uk}
\emailAdd{a.tolley@imperial.ac.uk}

\abstract{
Within the standard effective field theory of General Relativity, we show that the speed of gravitational waves deviates, ever so slightly, from luminality on cosmological and other spontaneously Lorentz--breaking backgrounds.
This effect results from loop contributions from massive fields of any spin, including Standard Model fields, or from tree level effects from massive higher spins $s \ge 2$. We show that for the choice of interaction signs implied by S--matrix and spectral density positivity bounds suggested by analyticity and causality, the speed of gravitational waves is in general superluminal at low--energies on NEC preserving backgrounds, meaning gravitational waves travel faster than allowed by the metric to which photons and Standard Model fields are minimally coupled.  We show that departure of the speed from unity increases in the IR and argue that the speed inevitably returns to luminal at high energies as required by Lorentz invariance. Performing a special tuning of the EFT so that renormalization sensitive curvature--squared terms are set to zero, we find that finite loop corrections from Standard Model fields still lead to an epoch dependent modification of the speed of gravitational waves which is determined by the precise field content of the lightest particles with masses larger than the Hubble parameter today. Depending on interpretation, such considerations could potentially have far--reaching implications on light scalar models,  such as axionic or fuzzy cold dark matter.
}

\maketitle


\section{Introduction}

In this new era of gravitational wave astronomy, it is especially important to understand how gravitational waves propagate. The recent simultaneous observation of gravitational waves  from the coalescence of two neutron stars, GW170817, together with its gamma--ray counterpart, GRB 170817A, has put the cleanest constraint on the propagation speed of gravitational waves relative to photons, \cite{TheLIGOScientific:2017qsa,Monitor:2017mdv,GBM:2017lvd}.\\

In {\it classical} General Relativity minimally coupled to matter, gravitational waves always travel luminally, as defined by the lightcones of the metric $g_{\mu\nu}$ with respect to which matter is coupled, by virtue of the equivalence principle. For instance, when considering  the propagation of linearized gravitational waves across some general curved background geometry, the background metric may always be put in a Riemann normal coordinate system where it is locally Minkowski in the vicinity of a spacetime point $x$, plus curvature corrections that grow away from $x$. Since the Einstein--Hilbert action is second order, modifications from the background curvature terms to the propagation of gravitational waves on this background can only arise as an effective mass term (simply from power counting derivatives), and never as corrections to the kinetic or gradient terms. For example, in FLRW spacetime, gravitational waves have an `effective mass' from the background expansion of order $H^2, \dot H$, in terms of the Hubble parameter $H$, but their sound speed defined by the ratio of kinetic to gradient terms is luminal\footnote{Canonically normalized tensor fluctuations in GR have quadratic action $S_{\rm GR}= \int \d \eta \d^3 x \frac{1}{2} \( h'^2-(\vec \nabla h)^2 + \frac{ a''}{a} h^2\)$. Despite the `effective mass' $-a''/a$ the actual mass is zero. In the well known case of the propagation of gravitational waves during inflation, this effective mass is negative, and drives an instability which generates long wavelength scale invariant tensor fluctuations, but the retarded propagator vanishes outside the lightcone defined by $c_s=1$. In what follows, in FLRW we define the speed via the lightcone of the effective metric on which modes propagate, i.e. via $c_s$ in the action $S_{hh}= \int \d \eta \d^3 x \frac{1}{2} \( h'^2-c_s^2 (\vec \nabla h)^2 - m^2_{\rm eff} h^2\)$, so the effective mass does not play a role in the definition of the speed. }. Hence it is the two derivative nature of Einstein's theory, together with diffeomorphism invariance, that guarantees luminality in General Relativity. \\

{\it Classical} General Relativity is not however the real world. At a minimum, gravitational effects generated from quantum loops of known particles, e.g. the electron, will already generate modifications to Einstein gravity which alter this process. Many of these effects are finite (meaning free from renormalization ambiguities) and calculable. These effects are of course highly suppressed, being induced by loops, but any potential departure of the speed of propagation of gravitational waves from luminality is itself significant, not least because it impacts our understanding of the causal structure of a given theory. Various proposed extensions to four dimensional General Relativity, such as extra dimension models, or string theory, will also induce modifications to the Einstein--Hilbert Lagrangian that can potentially change the above picture. \\

The general framework to account for such corrections is well understood and goes under the umbrella of `Effective Field Theory for Gravity' \cite{Donoghue:1994dn,Donoghue:1995cz,Burgess:2003jk,Donoghue:2012zc} (for a recent example of this methodology, see \cite{Ruhdorfer:2019qmk}). Historical issues with non--renormalizability and the artificial separation of quantum fields on curved spacetimes are replaced with the general effective field theory (EFT) framework that allows us, if desired, to simultaneously quantize matter and gravity despite the non--renormalizability of the Lagrangian. The price to pay is the need to introduce an infinite number of counterterms, but in practice at low energies only a finite number are ever relevant. The low energy effective theory is defined an effective Lagrangian valid below some scale $E\ll M$ which accounts for all tree and loop level corrections from particles of masses greater or equal to $M$, and loop processes of light fields at energies greater than $M$\footnote{Precisely how this is achieved depends on renormalization prescription, and we work with the most convenient which is dimensional regularization.}.
The starting point is then the Wilsonian action which includes all possible covariant operators build out of the Riemann tensor, its derivatives, and combinations of light fields and their derivatives.
For instance, assuming no other light fields than gravity, the leading corrections are powers of curvature and derivatives thereof, i.e. very schematically
\ba
S_{\rm EFT} &=& \int \d^4 x \sqrt{-g}  \left[ -\Lambda+ \frac{\mpl^2}{2} R + C_1 R^2 + C_2 R^2_{\mu \nu}+C_3 R^2_{\mu \nu \alpha \beta }
+ \sum_{n=1}^{\infty}  \frac{C_{n,p}}{\mpl^{2n}}\sum_{p=0}^{n} \nabla^{2 p} {\rm Riem}^{2+n-p}\nn\right] \,,
\ea
where by $\nabla^{2 p} {\rm Riem}^{2+n-p}$ we mean all possible scalar local operators constructed out of contractions of this number of powers of Riemann tensor and covariant derivatives.
The precise energy scale of suppression $C_{n,p}/\mpl^{2n}$ will depend on the origin of a given term, e.g. it will in general be different for interactions coming from tree level processes or from loops. In general there is not one such EFT, but a family of them depending on the choice of scale $M$ above which physics has been integrated out.\\

Even in the absence of matter, the speed of gravitational waves is modified by the addition of higher curvature terms, precisely because the earlier argument based on power counting of derivatives is no longer valid. Higher derivative curvature terms can, and do, modify the second order derivative terms in the equation for propagation when expanding around a background. Since gravitational waves are luminal in pure GR, the sign of higher curvature terms will typically lead the resulting corrections to make the waves either superluminal or subluminal.
Typically causality is imposed by demanding that the modified gravitational waves are subluminal (with respect to the lightcone defined by the metric $g_{\mu\nu}$, based on similar arguments for scalar fields \cite{Aharonov:1969vu,Adams:2006sv}, hence fixing the signs for the higher curvature coefficients. For instance for Ricci flat backgrounds this is done in \cite{Gruzinov:2006ie} for quartic curvature corrections of the type that arise in the low energy EFT from string theory. Within the context of the EFTs of inflation/dark energy, where matter sources the background, a potential modification to the speed of gravitational waves has been noted in \cite{Cheung:2007st,Gubitosi:2012hu,Gleyzes:2013ooa}. \\

More generally this procedure of demanding subluminality of all fluctuations is problematic because in a gravitational EFT the metric itself is ambiguous (see \cite{deRham:2014lqa} for related discussions). It is always possible to perform fields redefinitions, schematically of the form
\ba
\label{ambiguity}
g_{\mu \nu} \rightarrow g_{\mu\nu} + \sum_{p,n}\frac{\alpha_{n,p}}{\mpl^{2(n+ p)}}  \( \nabla^{2 p} {\rm Riem}^n \)_{\mu \nu}\,,
\ea
where $\( \nabla^{2 p} {\rm Riem}^n \)_{\mu \nu}$ is a tensor constructed out of $n$ contractions of Riemann and $2p$ covariant derivatives, which leave invariant the leading Einstein--Hilbert term. Those are consistent with the gravitational EFT, but modify the lightcone of the metric. In this way, in some cases, some spacetime with superluminal fluctuations may be rendered subluminal and vice versa\footnote{Although in general there is no universal procedure to render all fields (sub)luminal.}. A related effect known to occur at one loop is that the paths of massless particles of different spin do not receive the same amount of bending as they pass a massive object (e.g. the Sun) \cite{Bjerrum-Bohr:2014zsa,Bai:2016ivl,Chi:2019owc}, which means that despite being massless they effectively do not see the same metric which further confuses the question of how to describe the causal structure. \\

In response to this, more recent discussions of causality in the EFT context have focused on causality constraints implied by S--matrix analyticity. These have the virtue of being invariant under fields redefinitions and are in some sense true avatars of causality. One such idea is to demand causality {\it \`a la Wigner} by imposing positivity of the scattering time delay \cite{Camanho:2014apa,Afkhami-Jeddi:2018apj}. These arguments may plausibly apply for weakly coupled UV completions where the irrelevant operators in the EFT arise from tree level effects of high mass modes, but are already known to fail for QED where the corrections come from loops \cite{Hollowood:2015elj}. Another proposal is to use S--matrix positivity bounds \cite{Pham:1985cr,Ananthanarayan:1994hf,Adams:2006sv,deRham:2017avq,deRham:2017zjm}  which constrain the $2-2$ scattering amplitude of gravitons and other particles.  For example S--matrix positivity arguments have been applied to quartic curvature interactions in \cite{Bellazzini:2015cra}. Recent works have applied these ideas more specifically to the weak gravity conjecture, which focusses on the EFT for gravity and a $U(1)$ gauge field \cite{Bellazzini:2019xts,Mirbabayi:2019iae,Cheung:2018cwt}. In what follows we shall see that positivity bounds will provide very useful guidance on corrections to propagation speeds.
\\

Phenomenologically it is still pertinent to ask what is the speed of gravitational waves relative to the metric to which photons and the Standard Model fields couple minimally, and this is independent of field redefinitions. It is well known that the photon speed can be modified in a curved background due to loop corrections from charged particles, e.g. electrons, even leading to superluminal group velocities at low energies for certain backgrounds \cite{Drummond:1979pp,Lafrance:1994in}. The fact that this low energy superluminal group velocity is not in conflict with causality has been discussed extensively in a series of papers \cite{Hollowood:2007kt,Hollowood:2007ku,Hollowood:2008kq,Hollowood:2009qz,Hollowood:2010bd,Hollowood:2010xh,Hollowood:2011yh,Hollowood:2012as,Hollowood:2015elj}, which essentially identify the requirement that the front velocity as luminal as the key requirement for causality.  Apparent low energy violations of causality in, for instance scattering time delays, are absent in the UV theory \cite{Hollowood:2015elj}. In the EFT description these effects comes from non--minimal Riemann curvature coupling to the Maxwell field strength squared, specifically $\Delta {\cal L} \propto m_e^{-2}\sqrt{-g} R_{abcd} F^{ab} F^{cd}$, which would arise from electron loops. As such this effect only arises in the EFT defined below the scale of the electron, $\nabla \ll m_e$, or whatever charged particle has been integrated out. In order to have wider applicability to, in particular, the cosmological context, we shall assume matter (including photons), minimally couples to the metric, so that such equivalence principle violating curvature terms are not present. We shall rather focus on pure curvature interactions that can arise equally from integrating out charged and chargeless particles and are applicable for any matter (dark matter, Standard Model, inflaton etc.). In other words, photons will always be luminal, and the relevant question is what is the speed of gravitational waves as compared with a luminal photon, or at least the metric to which the photon is minimally coupled. \\

Our principle focus will be to ask what is the speed, by which we mean the speed defined by the effective lightcone of the low energy equations of motion (specified precisely in Section \ref{SpeedID}), of gravitational waves on a spacetime with a long range spontaneous breaking of Lorentz invariance. Thus we will not be interested in the rather special shock wave or asymptotically flat geometries considered for example in \cite{Camanho:2014apa,Hollowood:2015elj}, but in cases for which the departure of the speed of sound from unity can significantly build up over time to lead to clearly noticeable differences. The clearest example is FLRW spacetimes since they spontaneously break time diffeomorphisms, have a clearly identifiable sound speed, have sufficient symmetry to be simple such that gravitational modes decouple from matter at linear order, and have obvious phenomenological relevance. Nevertheless much of what we will discuss will be relevant to other more generic backgrounds. Crucially this means we do not consider vacuum spacetimes, but require some light fields to source the breaking. The inclusion of light fields in the EFT is useful since they themselves provide a clock, and their interactions with other matter can be used as we will see, to impose S--matrix analyticity and unitarity requirements.\\

\noindent Our main conclusions are:
\begin{itemize}

\item In the frame in which matter is minimally coupled, the leading EFT corrections with signs imposed by S--matrix locality, unitarity and analyticity (if the contribution from the massless graviton $t$--channel pole can be ignored) enforce that gravitational waves are superluminal for any matter satisfying the null energy condition (NEC).

\item On performing a field redefinition, this is equivalent at leading order (alone) to the generation of universal gravitationally induced matter interactions ($TT$ deformation) which ensure that standard matter fluctuations propagate slower than gravity.

\item The precise coefficient that determines the departure of the propagation speed from unity is connected with the elastic scattering amplitudes for matter fields, both those that drive the expansion and spectator fields.

\item If the leading order EFT corrections are set to zero, the next to leading order corrections that arise from loops give rise to an epoch and species dependent modification of the speed of gravitational waves. If the lightest particle with mass above the Hubble parameter has spin--0, the speed of gravity is superluminal throughout the whole standard cosmological history of the Universe.

\item Our results remain valid when considering purely quartic curvature corrections, such as those known to arise in the low energy string theory effective action, for which gravitational waves also travel superluminally on NEC preserving backgrounds.

\end{itemize}

Stated differently, given the assumed sign of the leading EFT coefficients (either as inferred from explicit integration of fields, or from positivity bounds or as implied from string theory), the lightcone inferred from the low energy sound speed of (minimally coupled) matter always lies inside the lightcone of gravitational waves and is never {\it exactly} at the same speed. The superluminality of the propagation of gravitational waves in the calculation in the original frame is, (far from being in conflict with), consistent with the requirements of causality implied by S--matrix analyticity. Arguably we must accept it as a price for the associated field redefinition ambiguity in the metric \eqref{ambiguity}, however as is already clear from the QED case \cite{Drummond:1979pp}, there are no field redefinitions which render all modes luminal or subluminal. We will further show that when additional non--minimal EFT interactions between gravity and the light fields are included, the same results hold, namely that positivity bounds (if valid) enforce the overall superluminality.\\

The rest of this paper is organized as follows: In Section~\ref{sec:GREFT} we start by reviewing what we need of the standard EFT of general relativity and how curvature corrections are generated in the low--energy EFT. We emphasize the role played by field redefinitions and how to take care of them.  We then explore the leading curvature--squared contributions to the low energy EFT for gravity in Section~\ref{sec:FLRW} and identify their effect on the speed of gravitational waves on FLRW and on static warped backgrounds.  We then discuss the implications of our findings within the context of standard causal and local UV completions in Section~\ref{sec:Super} and argue that such completions favour superluminal gravitational waves for NEC preserving backgrounds. In Section~\ref{sec:Rcubed} we explore the possibility of tuning the EFT so that the leading quadratic curvature corrections cancel, such that the dominant effect comes from higher (cubic and quartic) curvature corrections. In particular we show that the low--energy speed of gravitational wave depends on the field content of the high--energy completion, and particularly on the spin of the lightest massive particle that is integrated out to derive the low--energy Wilsonian action. We provide an outlook of our results in Section~\ref{sec:discussion}. Appendix~\ref{app1:1loop} derives the exact curvature--cubed operators in the one--loop effective action obtained from integrating out a massive scalar. Appendix~\ref{app2:FLRW} provides  the details for the derivation of the speed of gravitational waves on FLRW in the presence of curvature--cubed (dimension--6) operators. Finally Appendix~\ref{app:resumming} highlights subtleties in defining the retarded propagator perturbatively and justifies the approach we follow in identifying the speed.
\\

Throughout this paper we work in natural units where the speed of light in vacuum, without any quantum correction effects is $c=1$. We shall also, as is standard, slightly abuse the EFT operator counting terminology and refer to ${\rm Riemann}^n$ operators as dimension--$2n$ operators even though they include an infinite number of operators of various dimensions $\sum_k h^k (\p^2 h)^n/\mpl^{k+3n-4}+\cdots$.

\section{Effects of Heavy modes on Gravity at Low Energy}
\label{sec:GREFT}

Throughout this work, we consider gravity as a low--energy EFT and look at the effects that heavy fields minimally coupled to gravity have on the EFT. In other words we shall focus on the Wilsonian effective action for the light fields which shall include gravity as well as some light field that sources the background expansion (e.g. radiation, quintessence, inflaton) and look at the influence of those corrections that arise from integrating out massive fields for which the masses satisfy $M_i \gg H$. Our focus will be on identifying the speed of tensor gravitational waves on non--maximally symmetric backgrounds, and for most of Section~\ref{sec:FLRW} onwards we shall focus on cosmological backgrounds.
Working on FLRW has the advantage that at the linear level tensor fluctuations cleanly decouple from scalar and vector perturbations,
  the former of which is usually coupled to whatever matter drives the cosmological expansion. The FLRW symmetry also allows us to cleanly distinguish between the speed of gravitational waves, and the speed of matter perturbations. However we emphasize that our results hold more generically beyond FLRW as is for instance illustrated in Section~\ref{sec:warpedBackground} dealing with static warped geometries.

\subsection{Effective Field Theory for Gravity at Low--energy}

We shall have in mind two different scenarios:

\begin{itemize}

\item Tree level corrections to the EFT, whereby tree level effects of massive particles potentially generate higher curvature interactions

\item Loop level corrections to the EFT coming from integrating out standard matter (e.g. Standard Model fields) of any spin, including $s \le 1$.

\end{itemize}

The latter case is the most interesting since it does not require any assumptions about an unknown UV completion, but rather relies on the calculable gravitational effects of known particles, in particular from Standard Model particles.

\subsubsection*{Tree level interactions}

At tree--level, a massive field which is minimally coupled to gravity (i.e. without explicit curvature couplings) has no effect on the low--energy gravitational propagator if the spin of that field is less than two $s < 2 $. This can be seen straightforwardly as a consequence of the SVT decomposition. At tree level, we can work with the (partial) UV completion that includes the additional massive mode as part of the Lagrangian. If this mode has spin $s<2$ then it has no tensor component and so will at the level of quadratic fluctuations completely decouple from the gravitational tensor modes. Thus integrating out the massive mode at tree level will provide no contributions to the gravitational fluctuations. Stated differently, it is not possible to integrate out massive states with $s< 2$ at tree level and obtain higher curvature interactions that change the speed of gravitational waves. \\

The situation changes if we integrate out at tree level massive spins with $s \ge 2$. Any spin of $s \ge 2$ will by virtue of the helicity or SVT decomposition effectively contain spin $2$ states, or more precisely tensor modes. Even at quadratic order these states could mix with the usual massless graviton, generating modifications to its speed of propagation. We shall see explicit examples of this below. Obvious examples are string theory, where an infinite tower of massive spin states arise, or extra dimensional models where the massless graviton in higher dimensions can be viewed as a massless graviton in four, together with an infinite tower of massive spin 2 states. \\

Related arguments tell us that as soon as we allow for massive spin 2 states, in order to construct a weakly coupled UV completion of gravity we must necessarily include an infinite number of spin particles. Recent versions of these arguments have been given in \cite{Caron-Huot:2016icg,Camanho:2014apa,Afkhami-Jeddi:2018own} but they follow straightforwardly from the observation that the scattering amplitude of massive spin--2 particle violates the fixed $t$ Froissart bound by virtue of the $s^2$ growth of its $t$--channel pole, and in a weakly coupled UV completion this can only be turned around by an infinite number of powers of $s$ resumming into a softer behaviour, which necessitates an infinite number of spin states\footnote{The pole itself cannot be cancelled as its residue is positive by unitarity.} (e.g. see \cite{deRham:2018qqo}).

\subsubsection*{Loop level interactions}

At loop--level the situation is quite different. An internal loop, even of a particle of spin $s \le 1$,  effectively contains states of total angular momenta of arbitrary spin, as is implied by the partial wave expansion. As such loop corrections from standard matter can, and do, correct the propagation of gravitational waves. This effect is of course tiny, being loop suppressed, nevertheless it is finite (up to local counterterms), calculable, and under control from the EFT point of view. It is thus not necessary to know what the appropriate theory of quantum gravity is in order to determine the magnitude of this effect.  In what follows we will be integrating out heavy modes of mass $M$, with $H\ll M \ll \mpl$, where $H$ is the typical scale at which we are interested in probing our low--energy EFT for gravity (for instance $H$ is the typical scale of the curvature and we will consider modes with frequency $H\ll k \ll M$). We will only be integrating out loops of `matter fields' no gravitons in the loops. This is consistent with standard Wilsonian EFT, whereby we first integrate out massive states to construct the low energy EFT, from which light loops may be computed afterwards.
The former effects are captured by the Wilsonian effective action, the latter by the 1PI effective action. When focussing on the Wilsonian effective action, gravity, being a light field, is treated `classically' and so our results are largely independent of the precise details of quantum gravity and the UV completion of gravity at the Planck scale (or string scale). \\

To be more concrete, we consider gravity (standard GR) minimally coupled to light fields one or more of whom will be used to generate the cosmological backgrounds, and heavy fields of mass $M_i$ which define the UV completion. In the case of loop calculations we shall consider heavy fields of spin--$s=0, 1/2$ or $1$, but for tree level UV completions we have in mind any spin. We denote the massive spin fields generically by $\Phi$, and so schematically we have the action
\ba
\L_{\rm UV}=\sqrt{-g}\left[\frac{\mpl^2}{2}R+\L^{\rm (l.e.)}_\psi(g, \psi)+ \L^{\rm (h.e.)}(g, \Phi)+ {\cal L}_{\rm c.t.}
\right]\,,
\ea
where $\L^{\rm (l.e.)}_\psi$ is the low--energy Lagrangian for the low--energy fields (denoted generically as $\psi$) with masses $m_\psi \sim \mathcal{O}(H) \ll M$, (hence including massless modes), whose role will be to generate the cosmological background, while $\L^{\rm (h.e.)}(g, \Phi)$ represents the dynamics of the heavy fields, with masses $M_i\gtrsim M$.
We focus in what follows on minimal couplings between the both the light and heavy fields and gravity, meaning that the fields $\Phi$ and $\psi$ do not directly couple to the curvature below the Planck scale\footnote{We would of course expect the EFT for gravity to include operators that mix the Riemann curvature and the other fields through Planck scale suppressed terms. Such types of interactions are considered in Section~\ref{dim6lightfields}.}. \\

At loop--level, it is well--known that integrating out any massive field $\Phi$ would lead to divergent contributions to the cosmological constant, as well as to $R$ and curvature--squared terms $R^2$, $R_{\mu\nu}^2$ and $ R_{\mu\nu\rho\sigma}^2$. In the EFT context, in order to deal with these divergences we must add $\sqrt{-g}$, $\sqrt{-g}R$ and $\sqrt{-g}R^2$, $\sqrt{-g}R_{\mu\nu}^2$, $\sqrt{-g}R_{\mu\nu\rho\sigma}^2$ counterterms. Hence we must include in the UV action
\be
 {\cal L}_{\rm c.t.} = - \Lambda_{\rm UV} + \frac{1}{2} \mpl^2 a^{\rm UV} R + C_1^{\rm UV} R^2 + C_2^{\rm UV}R_{\mu\nu}^2+C_3^{\rm UV}R_{\mu\nu\rho\sigma}^2 +\dots\, ,
\ee
in addition to any other matter counterterms. The first two terms are just a redefinition of the cosmological constant and Planck mass and may be ignored in what follows as their consequences are straightforward. The latter terms are as we will see nontrivial, and directly affect the speed of propagation of gravitational waves.

\subsubsection*{Wilsonian Effective Action}

To construct the low energy Wilsonian effective action, we integrate out (both at tree and loop level) the heavy modes in the schematic manner
\be
e^{i S_{\rm IR}(g,\psi)} = \int D \Phi \, e^{i S_{\rm UV}(g,\psi,\Phi)} \, .
\ee
The resulting low energy effective theory will take the form
\ba
\L_{\rm IR}=\sqrt{-g}\left[-\Lambda^{\rm IR} +\frac{\mpl^2}{2}R+\L^{\rm (l.e.)}_{\psi}(g, \psi)+ C_1^{\rm IR} R^2 + C_2^{\rm IR}R_{\mu\nu}^2+C_3^{\rm IR}R_{\mu\nu\rho\sigma}^2+\dots
\right]\,.
\ea
More generally $\L^{\rm (l.e.)}_{\psi}(g, \psi)$ will also receive corrections if the light field couples to the heavy fields integrated out. An obvious and well known example is if the light field is the photon, integrating out charged particles will result at leading order in addition to the Euler--Heisenberg terms, the $R F F$ interactions considered in \cite{Drummond:1979pp} (see also \cite{Lafrance:1994in,Chu:2010tc}). In the interests of simplicity we will neglect these corrections for now, but consider examples of them in Section~\ref{dim6lightfields}. \\

The IR coefficients that enter the low energy EFT will differ from their UV values by virtue of both loop and tree effects
\be
C_{1,2,3}^{\rm IR} = C_{1,2,3}^{\rm UV} + \Delta C_{1,2,3}  \, .
\ee
The natural scale of the corrections $\Delta C_{1,2,3} $ is of order the number of fields integrated out $N$, $\Delta C_{1,2,3} \sim N$. In what follows we will see that positivity bounds generically imply that two specific combinations of these coefficients satisfy
\ba
&& \Delta C_{W^2}=\frac{1}{2} \Delta C_{2} +2 \Delta C_{3}>0 \, , \\
&&  \Delta C_{R^2} =\Delta C_1+\frac{1}{3}\Delta C_2 +\frac{1}{3}\Delta C_3>0  \, .
\ea
Indeed, we shall further argue, provided we may apply positivity bounds even in the presence of a massless graviton $t$--channel poles, as for example recently argued in \cite{Bellazzini:2019xts},  that\footnote{It is worth noting that this is the opposite sign to what is required for the quadratic gravity scenario \cite{Stelle:1976gc,Donoghue:2018izj,Donoghue:2019ecz,Lehners:2019ibe}. However it is expected that these models will have different causality and analyticity structure \cite{Coleman:1969xz,Donoghue:2018izj} and hence the usual positivity bounds are unlikely to apply.}
\be
C_{W^2}^{\rm IR}=\frac{1}{2} C_{2}^{\rm IR} +2 C_{3}^{\rm IR} > 0\, .
\ee
It is of course not possible in the EFT context to fix the precise values of $C_{1,2,3}^{\rm UV} $ or $C_{1,2,3}^{\rm IR}$  in the absence of an explicit matching calculation onto a UV completion. Hence we are instructed to compare them with observations. Precisely one such observation which is at least in principle possible to measure is the speed of gravitational waves relative to that of light. We shall begin in Section~\ref{Rsquared} by focussing on the case where these terms are present. In Section~\ref{sec:R3} we shall set them to zero and focus on the finite $R^3$ terms that arise from integrating out matter loops.

\subsubsection*{Inclusion of Light Loops}

As we have discussed the Wilsonian effective action $\L_{\rm IR}$ includes loops from heavy fields but not from light fields. As such it is local, and is the typical starting point for cosmological and phenomenological analyses. It is interesting to ask what would happen if we integrated out the light fields, in particular the massless graviton and photon. In this case we should be working with the 1PI effective action which is nonlocal and difficult to deal with. There is considerable work on this in the literature \cite{Barvinsky:1985an,Barvinsky:1995jv,Barvinsky:1994ic,Avramidi:1990je,Avramidi:1990ap,Barvinsky:1993en,Barvinsky:1994hw,Vilkovisky:2007ny,Codello:2012kq} and results are often presented in terms of a curvature expansion. At the level of curvature--squared terms, the contributions from loops of massless (or light) fields may be modelled by the following proxy effective action
\ba
\Delta L_{\rm light-loops} & = &\hat C_1^{\rm IR} R \ln\(\frac{-\Box-i \epsilon}{\mu_1^2} \) R + \hat C_2^{\rm IR}R_{\mu\nu} \ln\(\frac{-\Box-i \epsilon}{\mu_2^2} \) R^{\mu\nu}\\
&+& \hat C_3^{\rm IR}R_{\mu\nu\rho\sigma} \ln\(\frac{-\Box-i \epsilon}{\mu_3^2} \)R^{\mu\nu\rho\sigma} \, .\nn
\ea
This has been used in the cosmological context in \cite{Donoghue:2014yha}. It is clear that due to the logarithm, the massless loops can dominate over the heavy loop contributions, in particular in the IR. However, if as we will assume, the number of heavy fields is much greater than the number of massless or light fields, then we expect $C_i \gg \hat C_i$ and so it will be sufficient to focus on the heavy contributions.

\subsection{A Word of Caution on Field Redefinitions}

As is well--known, the $R^2$ and $R_{\mu\nu}^2$ interactions are redundant operators and are therefore removable with field redefinitions. Since the S--matrix is invariant under field redefinitions, it seems appropriate to ignore these contributions. This would be true for pure gravity, but when gravity is coupled to matter, all the field redefinition does is shift the same effect into another operator that arises at the same scale, specifically into a pure matter contribution that produces the same effect. In general field redefinitions of the metric change the `speed' of propagation by virtue of modifying the background metric with respect to which the speed is identified. However field redefinitions do not change the relative speed. For instance, if gravitational waves travel faster than photons in one `field frame', they do so in all 'field frames'. That is in the cosmological context
\be
\label{eq:ratiospeed}
\frac{c_s^2(\rm tensor)}{c_s^2(\rm matter)} \quad \text{is invariant under field redefinitions.}
\ee
The relative causal structure is kept in tact \cite{deRham:2014lqa}. Thus the question of whether gravitational waves are superluminal or subluminal with respect to light is a frame independent question.  \\

For backgrounds with FLRW symmetry, it is always technically possible to perform a field redefinition that renders the gravitational waves luminal. This is largely a triviality, due to the symmetry, the difference between the metric which matter couples to and the metric on which gravitational waves propagate is just a rescaling of the time component of the metric combined with an overall conformal factor (see e.g. \cite{Creminelli:2014wna}). Given whatever field is used to spontaneously break time diffeomorphism, $\psi$, we can always perform a field redefinition in the manner (as an example)
\be
g_{\mu\nu} \rightarrow A(\psi) g_{\mu\nu} + B(\psi) \nabla_{\mu} \psi \nabla_{\mu} \psi \, ,
\ee
and engineer the functions $A$ and $B$ so that for a given background the metric travels luminally. However, there is in general no single {\bf local} and {\bf covariant} background field redefinition that would render gravitational modes luminal around all backgrounds and so this procedure while comforting is also misleading. At one--loop level and higher order, we find Riemann cubed terms in the effective action generated from loops \eqref{eq:1LEFT}, part of which are Weyl cubed terms. These terms cannot be removed with a local field redefinition since they are not proportional to the leading equations of motion. Although these terms do not contribute on FLRW backgrounds, for backgrounds with less symmetry they do change the speed of gravitational waves and yet there is clearly no local field redefinition that removes them. \\

In this work we shall mainly focus on dimension--4 $R^2$ and dimension--6 ($R^3$ and $R\nabla^2 R$) curvature operators, as well as specific dimension--8 $R^4$ curvature operators in Section~\ref{sec:R4}. Dimension--4 curvature operators are naturally the leading contributions, but if those vanish (as will be considered in Section~\ref{sec:Rcubed}) the dimension--6 curvature operators are then the leading contributions.

\subsubsection*{Taking Care of the Dimension--4 Curvature Operators}

In discussing EFT descriptions of gravitational waves from mergers, in \cite{Endlich:2017tqa} it was argued that the dimension--4 could be removed via a field redefinition and are hence irrelevant for the low-energy EFT relevant for GWs. While it is true that such operators could be removed via field redefinitions, this would then affect $\L^{\rm (l.e.)}_\psi(g, \psi)$ and lead to non--minimal couplings with low--energy matter fields (and in particular photons), hence leading to a non--standard light--cone for light and other light particles. This effect is less important for the analysis there, but is crucially important for cosmological analyses. Here we largely insist on keeping a minimal coupling for the low--energy matter fields and hence avoid performing such field redefinitions accept where it useful to give an alternative explanation of the same phenomena and in deriving positivity bounds.

\subsection{Relevance of the Dimension--6 Curvature Operators}

As for the dimension--6 curvature operators, a subset of them are not removable by field redefinitions (namely the Weyl cubed terms). We shall consider those that arise in the specific computations of loops from particles of spin $s \le 1$ in Section~\ref{sec:Rcubed}. It was argued in \cite{Endlich:2017tqa} in the application gravitational waves from mergers, that those terms should be suppressed, and one should focus instead on dimension--8 operators. This argument was on the grounds that for weakly coupled UV completion these terms would come it a scale $\mpl^2 {\rm Riemann}^3/M^4$ where $M$ is the scale of particles that have been integrated out. In order for the ${\rm Riemann}^3$ term to have an interesting effect for gravitational wave astronomy, the scale $M$ would have to be taken so low that we would have observed the effect of the associated additional gravitationally coupled states that arise at the scale $M$. However, if the dimension--6 operators are suppressed, the dimension--8 operators must be further suppressed since it equally holds for dimension--8 operators that if they arise in a weakly coupled UV completion, they do so in the manner $\mpl^2 {\rm Riemann}^4/M^6$, then the UV completion would need an infinite number of states of spin $s>2$ arising at the same low scale $M$. This is transparent by their influence on the speed of propagation of gravitational waves, effects which could not arise at tree level in a theory of massless spin 2 and massless/massive spin $s<2$. Thus it does not make sense to argue on phenomenological grounds that the dimension--8 operators are larger than the dimension--6 ones. Lower dimension operators are always more significant, unless they are suppressed by a symmetry, which is not the case here. \\

In this work we shall not assume any such preconditions, and will consider the (albeit small) effect of all dimension--4, --6 operators  from loop corrections where they rather arise at the respective scales ${\rm Riemann}^2$,  $ {\rm Riemann}^3/M^2$ from integrating out particles of any spin, including Standard Model particles of mass $M$. By relinquishing ourselves from the constraints of purely tree level effects, we may consider lower mass scales $M$ without being ruled out by other gravitational constraints. We refer to Appendix~\ref{app1:1loop} for details on how integrating  out a massive scalar field, leads to specific ${\rm Riemann}^2$ and  $ {\rm Riemann}^3/M^2$ operators. The contributions from integrating out  generic spin--$s<2$ fields can be found in \cite{Avramidi:1990je}. We will also consider those dimension--8 operators known to arise in the string effective action in from $\alpha'$ corrections in Section~\ref{sec:R4}.

\section{Tensor Modes on a Background, Leading EFT Corrections}
\label{sec:FLRW}

In what follows we shall remain agnostic on the precise low--energy field content that leads to the cosmological solution and only assume that it is as `standard' as possible, in particular we assume that the effective degrees of freedom relevant for the low--energy dynamics (and the cosmological background)  are of spin $s<2$ (in particular it excludes massive gravity \cite{deRham:2010kj,deRham:2014zqa}), and couple minimally to gravity\footnote{Conformal couplings to gravity can be dealt with by first diagonalizing and then working with the appropriate minimally coupled  fields.}.

\subsection{Identifying Speeds in an EFT Context}
\label{SpeedID}

\subsubsection{Reorganizing EFT expansion}

Before proceeding to explicit calculations, it is worth discussing how we identify the speed of propagation in a time or space--dependent setting in which we are working, with a truncated EFT with higher derivative operators.
Throughout the following  discussion we mainly have FLRW in mind, although the reorganization of the EFT and the way we identify the speed is fully generalizable to any other type of background.

In an EFT higher derivative operators should be dealt with perturbatively, and we may only draw conclusions from them in the regime in which perturbation theory in these higher derivative operators is valid. For instance, working with the curvature--squared interactions introduces fourth order derivatives in the equations of motion. Directly perturbing will give an effective equation of motion for gravitational waves written in momentum space of the schematic form
\ba
\label{eq2}
 \(1+ \frac{\tilde b_2(\eta,k)}{\mpl^2}  \) \p_{\eta}^2 h_k(\eta)   + \(\tilde a_1(\eta) + \frac{\tilde b_1(\eta,k)}{\mpl^2} \)   \p_{\eta} h_k(\eta)+ \(\tilde a_0(\eta,k) + \frac{\tilde b_0(\eta,k)}{\mpl^2} \) h_k(\eta)  \nn\\
 +\frac{\tilde b_4(\eta)}{\mpl^2}  \p_{\eta}^4 h_k(\eta)+ \frac{\tilde b_3(\eta)}{\mpl^2}  \p_{\eta}^3 h_k(\eta) + {\cal O}(M^{-4},\mpl^{-2}M^{-2},\mpl^{-4}) \approx 0 \, .
\ea
To simplify the procedure we may  first perform a rescaling of field variables $h_k(\eta) \rightarrow \Omega(\eta) h_k(\eta) $ so that for the resulting equation the leading friction term $\tilde a_1(\eta) $ vanishes and the resulting momentum space equation is schematically of the  form
\ba
 \(1+ \frac{b_2(\eta,k)}{\mpl^2}  \) \p_{\eta}^2 h_k(\eta) +  \frac{b_1(\eta,k)}{\mpl^2}    \p_{\eta} h_k(\eta)  + \(a_0(\eta,k) + \frac{b_0(\eta,k)}{\mpl^2} \) h_k(\eta)  \nn\\
 +\frac{b_4(\eta)}{\mpl^2}  \p_{\eta}^4 h_k(\eta)+ \frac{b_3(\eta)}{\mpl^2}  \p_{\eta}^3 h_k(\eta) + {\cal O}(M^{-4},\mpl^{-2}M^{-2},\mpl^{-4}) \approx 0 \, .
\ea
In the limit $H \ll k/a \ll M$ we may use the WKB approximation to determine a dispersion relation which has 4 powers of frequency and hence twice as many solutions as that of a second order differential equation\footnote{The higher powers in the dispersion relation are always unphysical and simply signal the breakdown of perturbation theory when $k\sim  M$. At high energies the dynamics of the modes that have been integrated out should be accounted for. Note that the additional modes one would obtain in any truncated theory are not and should not be directly identified with the degrees of freedom present in the high energy theory \cite{deRham:2018dqm}.}. The additional solutions of this dispersion relation are the ghostly states that arise from the truncation and whose solutions should be ignored in the EFT context. What we are interested in are only the solutions that are {\it continuously connected} with the solutions that arise in the limit $\mpl \rightarrow 0$ for which the equation of motion is second order
\be
\p_{\eta}^2 h_k(\eta)=- a_0(\eta,k)  h_k(\eta)  +\mathcal{O}(\mpl^{-2})\, .
\ee
To identify these, we can either use the lower order equations and substitute them into the higher order ones, or more consistently at the level of the Lagrangian, perform field redefinitions to remove higher order time--derivatives.
Then  to the desired order, the equation of motion can be re--expressed as a second order in time derivatives equation of motion
\ba
\label{eq1}
 && \(1+ \frac{b_2(\eta,k)}{\mpl^2}  \) \p_{\eta}^2 h_k(\eta)   +  \frac{b_1(\eta,k)}{\mpl^2}    \p_{\eta} h_k(\eta) + \(a_0(\eta,k) + \frac{b_0(\eta,k)}{\mpl^2} \) h_k(\eta)    \nn \\
&&+ \frac{b_4(\eta)}{\mpl^2} \p_{\eta}^2  \left( - a_0(\eta,k)  h_k(\eta) \right) + \frac{b_3(\eta)}{\mpl^2} \p_{\eta} \left( - a_0(\eta,k)  h_k(\eta) \right) + {\cal O}(M^{-4},\mpl^{-2}M^{-2},\mpl^{-4}) \approx 0  \,.
\ea

\subsubsection{Identifying the Speed}
\label{sec:speedID}

We could at this stage use the WKB approximation to define an effective dispersion relation. Indeed we will in general only be interested in the effective speed of propagation in the region $H \ll k/a \ll M$ in which the EFT is under control and the modes are sufficiently subhorizon that the WKB approximation is valid and it is meaningful to talk about waves. However already for gravitational waves in GR, such an analysis would imply a superluminal group velocity when the effective mass is negative. For instance for GR, to implement the WKB approximation we begin with the ansatz
\be
h_k(\eta) = \frac{A}{\sqrt{ \omega(\eta, k)} }e^{i k z - i \int^{\eta} \d \eta' \, \omega(\eta', k)}\,,
\ee
which leads to the exact equation
\be
\label{WKB}
\omega^2(\eta,k) = k^2 - \frac{a''}{a} - \frac{1}{2} \( \frac{\omega''}{\omega} - \frac{3}{2} \frac{\omega'^2}{\omega^2} \)\, .
\ee
The WKB approximation amounts to solving this equation iteratively to any desired order. The leading iteration $\omega^2(\eta,k) = k^2 - a''/a$ would for example during inflation, where $- a''/a$ is negative, give superluminal $\d \omega/\d k$. This is clearly meaningless since an exact construction of the retarded propagator on FLRW shows that it only has support on and inside the lightcone \cite{Caldwell:1993xw}. The WKB approximation assumes $k^2 \gg - a''/a$ and hence is not accurate enough to account for the effective mass term in the exponent, all we can infer from it is $\omega^2 \approx k^2$ and that if this were the exact equation the front velocity $\lim_{k \rightarrow \infty} \omega/k$ would be luminal\footnote{The WKB approximation is still under control, it is just its interpretation that is failing. For example in the explicit case of gravitational waves in de Sitter $a=-1/(H \eta)$, the exact solutions are Hankel functions whose WKB form is $h \sim k^{-1/2} e^{i k z - i k \eta	} \( 1+\alpha_1 /(k \eta) + \sum_{n=2}^{\infty} \alpha_n /(k \eta)^2\)$. This follows from \eqref{WKB} by taking the leading iteration as $\omega=k$ and treating all high order terms perturbatively outside the exponent rather than inside it. In fact in this special case the series terminates $\alpha_n=0$ for $n \ge 2 $ and so this is the exact solution.
}.  What is relevant from the perspective causality, is neither the phase or group velocities which as we see are poorly defined in this time dependent setting, but the causal properties of the hyperbolic equations defining the retarded Green's functions. This is entirely determined by the lightcones of the hyperbolic metric defining the equation. \\

With this in mind we reorganize the EFT expansion in a manner suitable to determine the retarded propagator perturbatively, as a second order in time hyperbolic system plus (perturbative) higher spatial derivative corrections. In doing so it is worth emphasizing that in general the effective friction term, in the above example $(b_1-b_4 \p_\eta a_0-a_0 b_3)\p_\eta h_k$, is typically $k$--dependent and an additional rescaling  $h_k(\eta) \rightarrow \(1+\Omega(k,\eta)/\mpl^2 \)h_k(\eta) $ is helpful in order to remove any $k$--dependence in the friction term before determining the propagation speed. Once this is done, the equation of motion for tensors can be put in the form
\be
\label{eq3}
\p_{\eta}^2 h_k(\eta)  = - \sum_{n=0}^{\infty} \beta_n(\eta) k^{2n} h_k(\eta) \, ,
\ee
which is naturally reorganized as (to any desired order in the EFT expansion)
\be
\label{eq:iterating2}
\p_{\eta}^2 h_k(\eta) + \beta_1(\eta) k^2 h_k(\eta) + \beta_0(\eta) h_k(\eta) = - \sum_{n=2}^{\infty} \beta_n(\eta) k^{2n} h_k(\eta) \, .
\ee
The LHS defines a hyperbolic system with propagation speed $c_s^2 =  \beta_1(\eta)$ and effective mass $m^2_{\rm eff}=\beta_0(\eta)$. Temporarily ignoring the RHS, just as in the GR case, the presence or not of the effective mass is irrelevant to the causal structure of the retarded propagator. The latter is determined by the effective lightcone of the two derivative/$k^2$ terms. The full Green's function can be determined perturbatively by iterating the relation
\be
\label{eq:iterating1}
\p_{\eta}^2 G^k_{\rm ret}(\eta,\eta') + \beta_1(\eta) k^2 G^k_{\rm ret}(\eta,\eta')  + \beta_0(\eta) G^k_{\rm ret}(\eta,\eta')  = - \sum_{n=2}^{\infty} \beta_n(\eta) k^{2n} G^k_{\rm ret}(\eta,\eta')  + \delta(\eta-\eta') \, .
\ee
At any finite order in this expansion the causal structure of the Green's function will be determined by the zeroth order Green's function $G^k_{0 \, \rm ret}(\eta,\eta')$
\be
\p_{\eta}^2 G^k_{0 \, \rm ret}(\eta,\eta') + \beta_1(\eta) k^2 G^k_{0 \, \rm ret}(\eta,\eta') + \beta_0(\eta) G^k_{0 \, \rm ret}(\eta,\eta') =  \delta(\eta-\eta') \, ,
\ee
which has support on and inside the lightcones defined by the effective metric \cite{Caldwell:1993xw}
\be
\d \tilde s^2 = - c_s^2(\eta) \d \eta^2 + \d \vec x^2 \, .
\ee
Hence this defines what we mean by the low energy speed $c_s^2 =  \beta_1(\eta)$.  The above procedure may be easily generalized to any order in time derivatives and hence any order in the EFT expansion\footnote{Attempting to solve for the retarded Green's function perturbatively directly about the GR result would lead to a secular growth as explained in Appendix~\ref{app:resumming}.}. \\

In a fundamentally Lorentz invariant theory, all coefficients $\beta_n \rightarrow 0$ for $n\ge 2$ when the spontaneous breaking is removed, and similarly $\beta_1 \rightarrow 1$. Thus whilst the corrections to the sound speed from unity will be small, suppressed by at least one power of the maximal symmetry breaking,
eg. by one power of  $\dot H /\mpl^2$ in FLRW, the same will be true of all the coefficients $\beta_n$ for $n \ge 2$ which must similarly be suppressed by one power of $\dot H /\mpl^2$\footnote{Or higher derivatives of $H$, since on de Sitter $H$=constant there will be no modification to the speed by virtue of de Sitter invariance}. Hence, as long as $k/a$ is small in comparison to the momentum EFT cutoff, the dominant low energy modification to the propagation will be captured by $\beta_1-1$, that is
\be
|(\beta_1(\eta) -1) k^2| \gg \sum_{n=2}^{\infty} |\beta_n(\eta)| k^{2n}  \, .
\ee
The way in which we have defined the low energy speed is easily generalizable to perturbations around any spacetime and will always predict $c_s=1$ for pure GR minimally coupled to classical matter. The merit in its definition will be seen in that it is naturally connected with precise terms in scattering amplitudes that are potentially constrained by means of S-matrix positivity bounds as we discuss in Section~\ref{sec:Super}.

\subsection{Curvature--Squared Corrections on FLRW}
\label{Rsquared}

\subsubsection*{Dimension--4 Curvature Operators}

We begin in this section with the leading curvature corrections to our EFT, the $R^2$ corrections.  To reiterate, these may arising either from tree level or loop level effects of heavy fields, and are to this order
\be
{\cal L} = \sqrt{-g} \left[ \frac{\mpl^2}{2}R + {\cal L}^{\rm l.e.}(g,\psi)+ C_1 R^2 + C_2 R_{\mu\nu}^2 + C_3 R_{\mu\nu\rho\sigma}^2 +\cdots \right] \, ,
\ee
where the ellipses represent higher--order operators in the EFT expansion,
${\cal L}^{\rm l.e.}(g,\psi)$ is the Lagrangian for the low--energy fields $\psi$ which we assume here are all minimally coupled to gravity and do not include fields $\psi$ with spin 2 or more. Here and in what follows $C_i$ denotes the IR value $C^{\rm IR}_i$. In four dimensions $R_{\mu\nu\rho\sigma}^2$ can be written in terms of the Gauss--Bonnet term, which does not contribute to local dynamics, plus the remaining curvature--squared terms, and so these coefficients are better written in terms of the coefficients of Gauss--Bonnet, Weyl squared and Ricci scalar squared
\be
C_1 R^2 + C_2 R_{\mu\nu}^2 + C_3 R_{\mu\nu\alpha\beta}^2 = C_{R^2} R^2 + C_{W^2} W_{\mu\nu \alpha\beta}^2 + C_{\rm GB} \, {\rm GB} \, ,
\ee
where the Gauss--Bonnet term is
\ba
\label{eq:GB}
{\rm GB} = R_{\mu\nu\alpha\beta}^2- 4 R_{\mu \nu}^2 + R^2 \, .
\ea
The precise relations are $C_1=C_{R^2}+C_{\rm GB}+\frac{1}{3} C_{W^2}$, $C_2=-2 C_{W^2}-4C_{\rm GB}$, $C_3=C_{W^2}+C_{\rm GB}$.
Since the Gauss--Bonnet term is topological in four dimensions effectively for the rest of this section,  we shall be working with the leading EFT corrections to GR as follows
\be
\label{eq:R2Lag}
{\cal L} = \sqrt{-g} \left[ \frac{\mpl^2}{2}R + {\cal L}^{\rm l.e.}(g,\psi)+ C_{R^2} R^2 + C_{W^2} W_{\mu\nu \alpha\beta}^2 \right] \, .
\ee
As an example, we show in appendix~\ref{app1:1loop} how loops of a massive scalar field of mass $M$ leads to a contribution to such curvature--squared contributions with (see Eqn.~\eqref{eq:R2scalar})
\ba
C_{R^2}&=& \frac{5}{8}\frac{1}{16 \times 240 \pi^2}\log\(\frac{\Lambda^2}{M^2}\) \, ,\\
C_{W^2}&=& \frac{1}{4}\frac{1}{16 \times 240 \pi^2}\log\(\frac{\Lambda^2}{M^2}\)\,.
\ea

\subsubsection*{Tensor Modes on FLRW}

We now consider a FLRW background in conformal time $\eta$, with metric $\gamma\mn=a^2 \eta\mn$  and introduce the transverse and traceless tensor fluctuations $h_{ij}=\sum_\sigma h_\sigma \varepsilon^\sigma_{ij}$, where the sum is over the two polarizations $\sigma=+,\times$ and $\varepsilon^{+,\times}_{ij}$ represents the two polarization tensors. In what follows, we shall omit any mention of those two polarizations and simply denote the tensor modes as $h$, while it is of course understood that a sum over both polarizations is implicit. The tensor modes are   normalized so that the full metric is given by
\ba
\label{eq:hij}
g_{ij}=\gamma_{ij}+a h_{ij}\,.
\ea
We use the standard notation, $H=\dot a/a=a'/a^2$ is the Hubble parameter, with dots referring to derivatives with respect to physical time $t$, and primes with respect to conformal time $\eta$.  \\

\subsubsection*{Einstein--Hilbert}

The Einstein--Hilbert terms then leads to the standard canonical kinetic term for the tensor modes,
\ba
\L_{\rm EH}^{(hh)}=\frac{\mpl^2}{4}a^2   h  \(a^{-2}\Box_\eta -4 H^2-3 \dot H\) h \,,
\ea
where $\Box_\eta=-\p_{\eta}^2+\vec \nabla^2$ is the d'Alembertian on Minkowski spacetime, with $ \vec{\nabla}^2$ being the standard three--dimensional Cartesian Laplacian (that will be replaced by the momenta $-k^2$ below). \\

Assuming that all the couplings to gravity involved in $\L^{\rm (l.e.)}_\psi(g, \psi)$ are minimal and there are no fields $\psi$ with  spin $2$ or more, then the matter field Lagrangian leads to the following ``effective mass'' term for the tensor fluctuations on FLRW\footnote{This result follows for quite general Lagrangians, for instance for a single scalar $\psi$ it follows for all interactions of the form $P(X,\psi)=P((\p \psi)^2,\psi)$ as well as a generalized cubic Galileon $G((\p \psi)^2, \psi)\Box \psi$. However we do not consider other Horndeski operators as well as beyond Horndeski, as these involve non--minimal couplings to gravity and already induce a sound speed that differs from luminal at low--energy on cosmological backgrounds without even accounting for the effect of heavier modes.
},
\ba
\sqrt{-g}\L^{\rm (l.e.)}_\psi(g, \psi)^{(hh)}=\frac{\mpl^2}{2}a^2   \( 3H^2+2 \dot H\) h^2\,,
\ea
leading to the standard low--energy contribution
\ba
\label{eq:EHmi}
\L_{\rm EH, \psi}^{(hh)}=\frac{\mpl^2}{4}a^2   h  \(\frac 1{a^2}\Box_\eta +2 H^2+ \dot H\) h \,.
\ea

\subsubsection*{Curvature--Squared Contribution}

We shall now derive the equation of motion for tensor modes on including the $R^2$--operators which arise either as logarithmically running terms coming from matter loops, or may independently arise from tree level corrections from fields of spin $s\ge 2$.  The contributions from the $R^2$--operators are of the form
\ba
\label{eq:Ldim4a}
\L_{\rm dim-4}^{(hh)}=a^2\ h \ \hat{O}_{\rm dim-4} \ h\,,
\ea
where $\hat{O}_{\rm dim-{4}}$ is a 4th order operator given by
\ba
\label{eq:Odim41a}
\hat{O}_{\rm dim-4}=
\frac 1{a^4}g_{1}  \Box^2_\eta
+\frac 1{a^3}g_{3}  \Box_\eta  \p_{\eta}
+\frac 1{a^2}g_{4}  \Box_\eta
+\frac 1{a} g_{5}  \p_{\eta}
+\frac 1{a^2} g_{7}  \vec{\nabla}^2
+ g_{8} \,,
\ea
with the coefficients expressed as
\ba
\label{eq:g0}
g_{1} &=& C_{W^2}\\
g_{3} &=& 2 C_{W^2} H\\
g_{4} &=& 6 C_{R^2}(2H^2+\dot H)+6 C_{W^2} \dot H\\
g_{5} &=& -6 C_{R^2}(4H \dot H+\ddot H)-4C_{W^2} (H \dot H+\ddot H)\\
g_{7} &=&-4 C_{W^2} \dot H\\
g_{8} &=& 6 C_{R^2} \(4H^4-10H^2\dot H-8 \dot H{}^2-11 H \ddot H-2 \dddot H\)\\
&-& C_{W^2}\(2H^2 \dot H+\dot H{}^2+3 H \ddot H+\dddot H\)\nn
\,.
\ea
Working perturbatively in the dimension--4 curvature operators, following the approach discussed in Section~\ref{SpeedID}, we may substitute the relation for $\Box_\eta h$ in terms of $h$ as derived from \eqref{eq:EHmi},
\ba
\Box_\eta h = a^2(-2H^2-\dot H)h\,.
\ea
This perturbative substitution can be performed on the first three terms of the Operator $\hat{O}_{\rm dim-4}$ defined in \eqref{eq:Odim41a} so that only the last three terms remains with slightly altered coefficients,
\ba
\label{eq:Odim42a}
\hat{O}_{\rm dim-4}=
\frac 1{a} \tilde g_{5} \p_{\eta}
+\frac 1{a^2} g_{7} \vec{\nabla}^2
+ \tilde g_{8}\,.
\ea
The expressions of $\tilde g_{5,8}$ is irrelevant for the rest of the discussion but we include them for completeness,
\ba
\label{eq:gt5}
\tilde g_{5}&=&g_5+2 g_1(4H^3+6H \dot H+\ddot H)+g_3(-2H^2-\dot H)\\
\tilde g_{8}&=&g_8+g_1 (16H^4+34H^2\dot H+9H \ddot H+\dddot H+7(\dot H)^2)\\
&+&g_3(-4H^3-6 H \dot H-\ddot H)
+g_4(-2H^2-\dot H)\nn\,.
\label{eq:gt8}
\ea
At this stage we see directly that to this order, the modified equation of motion for the tensor modes on FLRW is
\ba
\frac{1}{a^2}\left[- \p^2_\eta + \(1- \frac{16 C_{W^2}\dot H}{\mpl^2}\)\nabla^2+ \frac{4a \tilde g_5}{\mpl^2} \p_{\eta}\right]h=m_0^2 h\,.
\ea
The friction term can easily be taken care of by performing a rescaling of the field which will keep the second space and time derivatives unaffected and simply modify the effective mass term by order $H^2/\mpl^2$ corrections. As a result we can directly read off the effective low--energy sound speed which as we see gets affected by the Weyl--term (and solely the Weyl term) on this spontaneously Lorentz breaking background,
\ba
\label{speedgravity1}
c_s^2=1+\frac{16 C_{W^2} (-\dot H)}{\mpl^2} +\mathcal{O}\(\frac{H^4}{\mpl^4}\)\,.
\ea
Interestingly, we see that the effective sound speed is superluminal on a null--energy condition satisfying background $\dot H<0$ as soon as $C_{W^2}>0$. At this stage we may be inclined to conclude that $C_{W^2}$ ought to be negative for any consistent (causal)  UV completion, however this conclusion would be wrong, or at least highly premature, as we will argue in what follows (see Section~\ref{sec:Super}). \\

In a weakly coupled UV completion, the natural scale for $C_W$ is $\mpl^2/\Lambda^2$ where $\Lambda$ is the scale of new tree level physics. Hence the correction to the sound speed may be as large as $\sim |\dot H|/\Lambda^2$. This is particularly interesting in the case of inflationary models where the hierarchy between $ |\dot H|^{1/2}$ and the scale of new physics $\Lambda$ is not necessarily large.

\subsection{Static Warped Geometries}
\label{sec:warpedBackground}

Although our main focus is on cosmological spacetimes, it is worth noting that the above analysis trivially generalizes to static warped geometries with $ISO(1,2)$ symmetry. By analytic continuation we can equally well consider metrics with non--trivial dependence on only one space dimension, e.g. $y$ and associated matter profiles
\be
\d s^2 = a(y)^2 ( \d y^2  - \d t^2 + \d x^2 + \d z^2) + a(y) h_{ab} \d x^a \d x^b\,,
\ee
where $h_{ab}$, transverse and traceless with respect to the $(t,x,z)$ subspace. Due to the fact that these solutions have the same amount of symmetry as the FLRW solutions, the equivalently defined tensor modes decouple from the matter degrees of freedom which source the background $y$ dependence.  Repeating the previous analysis, we find similarly a fourth order differential equation which may be reorganized into a second order differential equation with associated propagation speed
\ba
\label{eq:speedwarped}
c_s^2(y)=1+\frac{16 C_{W^2} (-H_y'(y))}{\mpl^2} +\mathcal{O}\(\frac{H^4}{\mpl^4}\)\,.
\ea
where $H_y(y)= \d \ln a(y)/\d y $ and $H_y'(y) = \frac{\d  H_y(y)}{a(y) \d y}$. As in the cosmological case, for matter satisfying the null energy condition we have
\be
H_y'(y)  < 0 \, ,
\ee
and so we again conclude that if $C_{W^2}>0$ then the tensor modes propagate superluminally. This is consistent with the arguments given below in Section~\ref{sec:Super} which apply for any geometry.

\subsection{Sound Speed Frequency Dependence}

\label{sec:freq}

We have seen that the speed of gravitational waves are superluminal in the low energy region for $C_{W^2}>0$. Since this calculation is performed in an EFT context, we can only trust the calculation of $c_s^2$ up to and including $1/\mpl^2$ corrections without including higher order operators in the EFT Lagrangian. Nevertheless it is instructive to see what happens if we temporarily assume that the $R^2$ and $W^2$ terms define the exact Lagrangian and compute the speed to higher order. The next order correction takes the form
\ba
\omega^2 \approx \left( 1+\frac{16 C_{W^2} (-\dot H)}{\mpl^2}
- 1024 C_{W^2}^3 \left( \frac{3{\dot H}^2-6 H^2 \dot H+5 H \ddot H-\dddot H}{\mpl^4}\right) \frac{k^2}{a^2 \mpl^2} + {\cal O}\(\frac{H^4 k^4}{\mpl^8}\) \right) k^2\, .\qquad
\ea
We see that at higher momenta, the departure of the speed from unity is reduced regardless of the sign of $C_{W^2}$. Indeed, determining the exact form of the dispersion relation shows that the speed of sound always asymptotes to unity as $k \rightarrow \infty$. This is simply because, at high energies in the truncated Lagrangian, the $W^2$ terms dominate the dispersion relation, the leading term is a Lorentz invariant $\Box^2$ operator. We stress again, that we cannot trust this calculation in the EFT context since other operators will kick on, however it is indicative of a general expectation that even on a background which spontaneously break Lorentz invariance, for momenta much larger than the scales of the background it will always be the leading Lorentz invariant operators which dominate and guarantee a luminal front velocity
\be
\lim_{k \rightarrow \infty} c_s^2(k) =1 \, .
\ee
Indeed taking seriously the fourth order equation inferred from \eqref{eq:Ldim4a} and \eqref{eq:EHmi}, it is helpful to note that to this order the effective action can be rewritten as
\be
S^{(hh)} = \int \d^4 x \, \left[  \frac{g_1}{a^2 }  (\Box_{\eta} h)^2 + \( \frac{\mpl^2}{4}+ g_4\) (\dot h^2 - (\vec \nabla h)^2 )-g_7 (\vec \nabla h)^2 +a^2 D(\eta) h^2 \right] \, .
\ee
This can be rewritten as a second order system by introducing an auxiliary variable $\Psi$
\be
S^{(hh)} = \int \d^4 x \, \left[   \Psi \Box_{\eta} h  - \frac{1}{4} a^2 g_1^{-1} \Psi^2 +  \( \frac{\mpl^2}{4}+ g_4\)  (\dot h^2 - (\nabla h)^2 )- g_7 (\vec \nabla h)^2 + a^2 D(\eta)  h^2 \right] \, .
\ee
Performing a standard WKB approximation with ansatz
\be
h = h_0(\eta,k) e^{-i W(\eta,k)} \, ,\quad
\Psi = \Psi_0(\eta,k) e^{-i W(\eta,k)}\,,
\ee
with  $e^{-i W(\eta,k)}$ varying rapidly in time and $h_0(\eta,k),\Psi_0(\eta,k) $ slowly in time,
we obtain at leading order in $k \gg a H$ the approximate equations
\ba
&& (W'^2-k^2) h_0 \approx \frac{1}{2} a^2 g_1^{-1} \Psi_0 \, ,\\
&& (W'^2-k^2) \Psi_0 +2  \( \frac{\mpl^2}{4}+ g_4\)  (W'^2-k^2) h_0 -2 k^2 g_7 h_0 \approx 0 \, .
\ea
Combining together the dispersion relation can be determined from
\be
\( \frac{\mpl^2}{4}+ g_4\) \( \frac{1}{a^2} W'^2 - \frac{k^2}{a^2}  \) + g_1 \( \frac{1}{a^2} W'^2 - \frac{k^2}{a^2}  \) ^2     - \frac{k^2}{a^2} g_7 \approx 0 \, ,
\ee
which has the exact solution (taking only that solution which is continuously connected with the usual GR solution)
\be
\omega^2 = W'^2 = k^2 + a^2 \frac{-4 g_4-\mpl^2+ 4 \sqrt{\frac{4 g_1 g_7 k^2}{a^2}+ (g_4+ \frac{\mpl^2}{4})^2} }{8g_1}
\ee
Interestingly this solution is always real, meaning no decay, regardless of the sign of $C_{W^2}$ (i.e. $g_1$) as long as the null energy condition is satisfied $\dot H<0$, has the desired low energy behaviour, and asymptotes to $\omega^2=W'^2 \rightarrow k^2$ at high energies. \\

We have avoided performing a WKB analysis of the fourth order equation in order to determine the speed of propagation in the previous sections since for these higher order derivative systems they generally do not give an accurate determination of the low energy speed. In particular, we see from performing a Taylor expansion of this approximation, the order $k^4$ term is incorrect. It is however correct in the higher momentum limit if the equation were taken as exact.

\section{Superluminality \& Causal UV Completions}
\label{sec:Super}

On NEC preserving backgrounds, we have seen that gravitational waves have superluminal low energy speeds if the coefficient $C_{W^2}$ of the Weyl--squared operator in the EFT of gravity is positive. If one were to jump to conclusions at this stage, one may be inclined in arguing that consistency of the EFT requires setting $C_{W^2}$ to be negative (this is indeed the logic followed in much of the standard literature). However as we shall argue in this section, this conclusion is premature and likely erroneous. Indeed as already argued earlier, contributions to the coefficient $C_{W^2}$ from tree--level massive spin--$s\ge 2$ fields, or from loops of massive spin--$s<2$ fields lead to a positive contribution to $C_{W^2}$. In what follows we shall show that a positive sign for $C_{W^2}$ typically follows from standard positivity bound arguments (if applicable to gravity) and ensures subluminality in the matter sector.

\subsection{Gravity versus Matter light cones and Null--Energy Condition}

\label{lightcone}

\subsubsection*{Matter Frame}

Both for NEC preserving FLRW backgrounds and for NEC preserving static warped geometries, it was shown in \eqref{speedgravity1}  and \eqref{eq:speedwarped} that the effective low energy sound speed of gravitational waves is (ever so slightly) superluminal as soon as $C_{W^2}>0$. This result can actually be derived very simply by recognizing that it is entirely a consequence of the field redefinition between the metric frames in which matter minimally couples and that in which gravity minimally couples, following from \eqref{fieldredef}. To see this explicitly, we can start with our EFT Lagrangian for GR including the leading order curvature corrections
as given in \eqref{eq:R2Lag}
\ba
\label{eq:R2Lag2}
{\cal L} &=& \sqrt{-g} \left[ \frac{\mpl^2}{2}R + C_{R^2} R^2 + C_{W^2} W_{\mu\nu \alpha\beta}^2 +  C_{\rm GB} {\rm GB}+ \L_{\rm matter}(g,\psi)\right] \,, \\
&=& \sqrt{-g} \left[ \frac{\mpl^2}{2}R + ( C_{R^2}- \frac{2}{3} C_{W^2} ) R^2 + 2 C_{W^2} R_{\mu\nu }^2 +  (C_{W^2}+C_{\rm GB}) {\rm GB}+ \L_{\rm matter}(g,\psi)\right] \,,
\ea
where $\L_{\rm matter}(g,\psi)$ is the Lagrangian for the (low--energy) matter fields that we assume (for now) are minimally coupled to gravity. For definiteness, we refer to this frame as the frame in which matter is minimally coupled and denote the metric in this frame as $g_{\mu\nu}=g_{\mu\nu}^{\rm matter}$. The leading order equation of motion in this form is $G^{\rm matter}_{\mu\nu}= \mpl^{-2} T_{\mu\nu}$, where $T\mn$ is the stress--energy of the matter field
\ba
T\mn=-\frac{2}{\sqrt{-g}}\frac{1}{\delta g^{\mu\nu}}\(\sqrt{-g}\L_{\rm matter}(g,\psi)\) \, .
\ea

\subsubsection*{Tensor Frame}

Consider now the following field redefinition
\ba
g_{\mu\nu}^{\rm matter} = g_{\mu\nu}^{\rm tensor}+ \frac{1}{\mpl^2}\delta g_{\mu\nu} \, ,
\ea
for which the Lagrangian picks up a new interaction
\be
\Delta {\cal L}= -\frac{1}{2} \delta g_{\mu\nu} ( G^{\mu\nu} - \mpl^{-2} T^{\mu \nu}) +\frac{1}{\mpl^2}\mathcal{O}\(R R^{\mu \nu}\) \, .
\ee
If we make the choice that
\be
\label{fieldredef}
\delta g_{\mu\nu} = 4 C_{W^2}  \(G_{\mu\nu}+ \frac{1}{\mpl^2}T_{\mu\nu}\) -2 ( C_{R^2}- \frac{2}{3} C_{W^2} ) \(R -\frac{1}{\mpl^2} T\) g_{\mu\nu} \, ,
\ee
then the field redefined Lagrangian is\footnote{To this order this is similar to a four dimensional $T \bar T$ deformation, however the coincidence ends at this order \cite{Taylor:2018xcy}.}
\be
\label{eq:LTT}
{\cal L} = \sqrt{-g} \left[ \frac{\mpl^2}{2}R +  C_{\rm GB} {\rm GB}+ {\cal L}_{\rm matter} + \frac{2 C_{W^2}}{\mpl^4} T_{\mu\nu} T^{\mu\nu} +\frac{( C_{R^2}- \frac{2}{3} C_{W^2} )}{\mpl^4} T^2+ \cdots \right] \, ,
\ee
where again ellipses represent higher order curvature operators (e.g. of order $R^3/M^4$) and $g_{\mu\nu}$ is not the tensor (Einstein) frame metric $g_{\mu\nu}^{\rm tensor}$. At the order at which we are working, the dimension--4 curvature--squared interactions have now disappeared, other than the Gauss--Bonnet term which does not affect local physics, at the price of non--minimal interactions in the matter sector. Such types of operators were considered within the context of EFTs for cosmic acceleration \cite{Park:2010cw,Bloomfield:2011np}. It is clear that to this order, in this representation, gravitational waves will travel at the speed defined by the lightcones of the metric $g^{\rm tensor}_{\mu\nu}$, but light itself will no longer travel at this speed since Maxwell's equations are modified by the inclusion of higher order operators.

\subsubsection*{Gravitationally Induced Matter Interactions}

\begin{figure}[h]
  \centering
  \includegraphics[width=\textwidth]{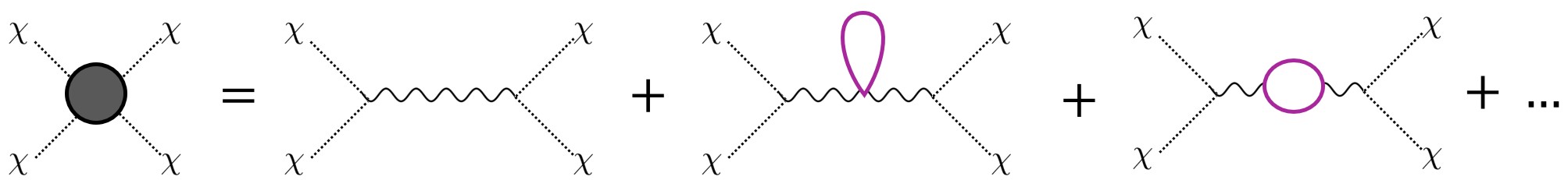}
  \caption{TT amplitude: Graviton mediated loop contributions to matter interactions. $\chi$ symbolizes a matter field present in the stress--energy tensor $T\mn$, a wiggly line is a graviton propagator and solid purple lines are the loops of heavy fields. }\label{Fig:loopTT}
\end{figure}

\begin{figure}[h]
  \centering
  \includegraphics[width=0.7\textwidth]{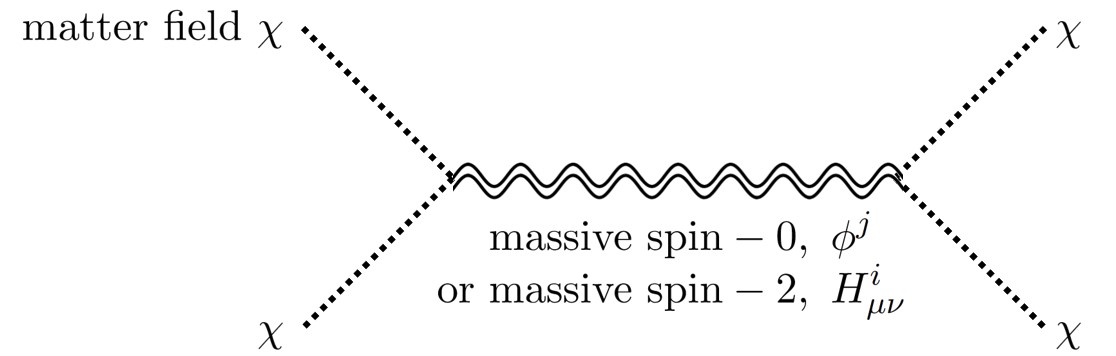}
  \caption{TT amplitude: Gravitational strength matter interactions arising from exchange of mass spin 0 and spin 2 states.}\label{Fig:WeakCoupling}
\end{figure}

This leading order `$TT$ deformation' of the matter Lagrangian in \eqref{eq:LTT} can be understood diagrammatically as arising from the process given in Figs.~\ref{Fig:loopTT} and \ref{Fig:WeakCoupling}. The diagram in Fig.~\ref{Fig:WeakCoupling} represents the tree level process whereby a massive heavy state of spin--2 or --0 is exchanged between the two stress energies. This corresponds to the explicit example given in Section~\ref{Weaklycoupled}. The diagram in Fig.~\ref{Fig:loopTT} describes a loop process from a heavy field mediated via tree level massless graviton exchange. This corresponds to the explicit example given in Section~\ref{sec:1lEFT}, at least after field redefinition. We stress again that while the perspective obtained by performing these field redefinitions is useful, at least at these low orders, once we consider higher order interactions, gravitational couplings arise (e.g. Riemann$^3$) which cannot be removed via local field redefinitions alone, it will not be possible to give such a simple effective description in terms of gravitationally induced matter interactions. Of course S-matrix elements are invariant under these field redefinitions, and the on-shell process described by Figs.~\ref{Fig:loopTT} and \ref{Fig:WeakCoupling} can be computed in any frame.

\subsubsection*{Connection with the NEC}

 After performing the field redefinition to remove the curvature--squared terms, we have the Einstein (or tensor) frame metric in which the gravitational tensor fluctuations are minimally coupled
\be
g_{\mu\nu}^{\rm tensor}=g_{\mu\nu}^{\rm matter} - \frac{4 C_{W^2}}{\mpl^2} (G^{\rm matter}_{\mu \nu} + \frac{1}{\mpl^2}T_{\mu\nu}) - \frac{2 C_{R^2}-\frac{4}{3} C_{W^2}}{\mpl^2} (-R^{\rm matter}+ \frac{1}{\mpl^2}T)  g_{\mu\nu}^{\rm matter} +\dots
\ee
Evaluating this on--shell, then to the same order we have
\ba
g_{\mu\nu}^{\rm tensor} &=&g_{\mu\nu}^{\rm matter} - \frac{8 C_{W^2}}{\mpl^2} G^{\rm matter}_{\mu \nu} - \frac{4C_{R^2}-\frac{8}{3} C_{W^2}}{\mpl^2} R^{\rm matter}  g_{\mu\nu}^{\rm matter} +\dots \\
&=& g_{\mu\nu}^{\rm matter} - \frac{8 C_{W^2}}{\mpl^4} T^{\rm matter}_{\mu \nu} + \frac{4C_{R^2}-\frac{8}{3} C_{W^2}}{\mpl^4} T^{\rm matter}  g_{\mu\nu}^{\rm matter} +\dots  \, .
\ea
For concreteness, we now focus on the FLRW metric considered in Section~\ref{Rsquared}, although the following argument clearly applied for any background (as for the example of the static warped geometry in Section~\ref{sec:warpedBackground}), not just FLRW. \\

When the matter metric has the FLRW form $g_{\mu\nu}^{\rm matter} \d x^{\mu} \d x^{\nu} = a(\eta)^2 ( - \d \eta^2+  \d \vec  x^2) $, then to this order the Einstein (tensor) frame metric is
\be
g_{\mu\nu}^{\rm tensor} \d x^{\mu} \d x^{\nu} = \Omega^2 a^2 ( - c_s^2\d \eta^2+ \d \vec  x^2) +\dots\,,
\ee
with $c_s^2= 1+16 C_{W^2} (-\dot H)/\mpl^2$ which is exactly the result obtained in \eqref{speedgravity1}. The conformal factor $\Omega^2$ is given by
\ba
\Omega^2=\(1 + \frac{8 C_{W^2}}{\mpl^2} (2 \dot H + 3H^2) -\frac{(24 C_{R^2}-16 C_{W^2})}{\mpl^2} (\dot H + 2H^2)\)\,.
\ea

One of the interesting features about the above results for the sound speed is that the correction is proportional to $\dot H$ and so changes sign if we consider a field theory with NEC violation.  A violation of the NEC, required to achieve $\varpi=p/\rho<-1$, typically leads to instabilities \cite{Creminelli:2006xe,Creminelli:2008wc} or even a breaking of the low--energy effective field theory \cite{deRham:2017aoj}, unless it is accompanied with superluminal modes in the sector responsible for the breaking of the NEC \cite{Dubovsky:2005xd}. Our findings naturally complement these results in the case where $C_{W^2}>0$. Indeed, for $C_{W^2}>0$, gravitational waves become subluminal as soon as $\dot H>0$, which, in the field frame in which gravity is luminal, is equivalent to the statement that the matter fluctuations become superluminal as soon as the NEC is violated (as soon as $\dot H>0$). \\

The previous argument works for any background given the field redefinition \eqref{fieldredef}.  If $n_{\mu}$ denotes a vector which is null with respect to the matter lightcone $g_{\mu\nu}n^{\mu} n^{\mu}=0$, then if matter satisfies the NEC $n^{\mu} n^{\nu}  T_{\mu\nu}>0$  we have,
\ba
\begin{cases}
  n^{\mu} n^{\nu} g_{\mu\nu}^{\rm tensor} < 0, & \mbox{if } C_{W^2}>0 \\
  n^{\mu} n^{\nu} g_{\mu\nu}^{\rm tensor} > 0, & \mbox{if }  C_{W^2}<0\,,
\end{cases}
\ea
meaning that the vector $n^{\nu} $ is timelike with respect to the gravitational wave lightcone if $C_{W^2}>0$ and spacelike if $C_{W^2}<0$. In the case where $C_{W^2}>0$ (as would for instance be the case if the curvature--squared corrections were solely arising from integrating out massive scalar fields or from a spectral representation discussion as will be provided below, see Section~\ref{sec:TTamplitude}, since the vector $n^{\nu} $ is timelike with respect to the gravitational wave lightcone, the matter lightcone always lies inside the gravity lightcone, for all NEC respecting matter, regardless of the choice of background. Violating the NEC or having $C_{W^2}<0$ automatically reverses this (arguably natural) order.

\subsection{Positivity Bounds for Light Fields}
\label{positivitysec}

From an EFT point of view the coefficients $C_{R^2}$ and $C_{W^2}$ are a priori undefined unless we match them with a UV completion as will performed shortly in Section~\ref{sec:TTamplitude}.
In the case where we are dealing solely with curvature--squared corrections in the gravitational EFT, a local field--redefinition is always possible so as to move the corrections into the matter sector as was performed in the previous subsection. Before considering a matching with a UV completion, we shall first consider how the standard positivity bounds from S--matrix analyticity, locality and unitarity may be used to constrain the sign of $C_{R^2}$ and $C_{W^2}$ of the resulting matter interactions, {\bf provided we argue or assume that the contribution of the graviton exchange $t$--channel pole can be neglected}. \\

\subsubsection*{Scaling Limit}

Assuming we are free to choose the coefficients in the EFT, at this point we can take a decoupling limit $\mpl \rightarrow 0 $ keeping $C_{R^2}/\mpl^2$ and $C_{W^2}/\mpl^2$ fixed. For instance in the case in which the $R^2$ terms arise from loop corrections from integrating out fields, we expect the $C$'s to scale with $N$, the number of fields. At the same time the Planck mass is related to the species scale as $\mpl^2 = N \Lambda_{\rm species}^2$ \cite{Dvali:2007hz}. Hence this decoupling limit is simply the limit $N\rightarrow \infty$, $\mpl \rightarrow \infty$ keeping $\Lambda_{\rm species}$ fixed. In this limit, gravity may be describe a linearized massless spin--2 on Minkowski coupled  to matter with Lagrangian
\be
\label{decoupling}
{\cal L } = \frac{1}{8}h^{ab} {\cal E}  h_{ab} + \frac{1}{2 \mpl} h^{ab} T_{ab}   +{\cal L}_{\rm matter}'+\dots
\ee
where $ {\cal E} =\Box+\dots$ is the Lichnerowitz operator, $T_{ab}$ is the stress energy of ${\cal L}_{\rm matter} $, and the last term is a modified matter Lagrangian
\be
{\cal L}_{\rm matter}'={\cal L}_{\rm matter} + \frac{2 C_{W^2}}{\mpl^4} T_{\mu\nu} T^{\mu\nu} +\frac{( C_{R^2}- \frac{2}{3} C_{W^2} )}{\mpl^{4}} T^2  + \dots \, .
\ee
Since this is now a local field theory living on Minkowski spacetime, we may ask what are the constraints on the coefficients $C_{R^2}$ and $C_{W^2}$ based on positivity bounds applied to the scattering of these light matter states, including these $N^{-1} \sim \mpl^{-2}$ corrections.

\subsubsection*{Light Scalar Fields}

For instance, suppose we consider matter to be a single (nearly) massless scalar field whose stress energy takes the form $T_{\mu\nu} = \partial_{\mu} \phi\partial_{\nu} \phi- \frac{1}{2} \eta_{\mu\nu} (\partial \phi)^2$. The effective Lagrangian for the scalar is then
\ba
{\cal L}_{\rm matter}'=-\frac{1}{2} (\partial \phi)^2  +\frac{( C_{R^2}+ \frac{4}{3} C_{W^2} )}{\mpl^4} (\partial \phi)^4  +\O(\mpl^{-4})\, .
\ea
At order $N^{-1} \sim \mpl^{-2}$, the tree $2-2$ scattering amplitude describing the process $\phi\phi \rightarrow \phi\phi$ will receive two types of contributions. Contact interactions which come from the $(\partial \phi)^4$ interactions and an $s$,$t$ and $u$ channel poles that come from the exchange of massless spin--2 graviton,
\be
\label{phiphiamp}
{\cal A}_{\phi\phi \rightarrow \phi\phi}(s,t) \sim \frac{1}{2\mpl^2} \( \frac{-t u}{-s} + \frac{-s t}{-u} + \frac{-s u}{-t} \) +2 \frac{( C_{R^2}+ \frac{4}{3} C_{W^2} )}{\mpl^4} (s^2+t^2+u^2) + {\cal O}(\mpl^{-4})
\ee
The direct application of forward limit positivity bounds \cite{Pham:1985cr,Ananthanarayan:1994hf,Adams:2006sv} is famously problematic due to the massless $t$--channel pole. Defining as $\mathcal{A}'$, the fixed $t$, $s$--channel pole subtracted amplitude, we have
\ba
\p_s^2 \mathcal{A}'(s,t)\sim -\frac{1}{\mpl^2}\frac{1}{t}+\frac{4}{\mpl^4}( C_{R^2}+ \frac{4}{3} C_{W^2} )+ {\cal O}(\mpl^{-4})\,.
\ea
The forward scattering limit of $\p_s^2 \mathcal{A}'(s,t)$ is dominated by the contribution from the $t$--channel pole and potentially bears no relevance for the sign of $C_{R^2}+ \frac{4}{3} C_{W^2}$. More importantly, the pole at $t=0$ prevents analytic continuation of the partial wave expansion from the physical region $t<0$ to $t>0$, precluding any statement of positivity even for small positive $t$\footnote{In the case of massive gravity, this problem is conveniently avoided since the pole is at $t=m^2$ and so one can analytically continue from $t<0$ to $m^2>t>0$. Extensions of positivity bounds to $t>0$ have recently been considered in \cite{deRham:2017avq,deRham:2017zjm} with particular application to massive gravity in \cite{deRham:2017xox,deRham:2018qqo}.}. \\

However a recent argument given in \cite{Bellazzini:2019xts} has suggested a potential solution. The idea is to regulate the IR divergence at $t=0$ by compactification to 3 dimensions and apply positivity bounds there.  Assuming the validity of this reasoning, it makes it possible to discard the contribution from the massless graviton $t$ channel pole. If correct, then in the present context the forward limit positivity bounds \cite{Pham:1985cr,Ananthanarayan:1994hf,Adams:2006sv} applied to pole subtracted amplitude impose
\be
 C_{R^2}+ \frac{4}{3} C_{W^2}  > 0 \, .
\ee
In the next section we shall argue for this positivity in a different manner which is consistent with this ability to ignore the $t$--channel pole.

\subsubsection*{Electromagnetism}

Similarly, taking the example of the matter being electromagnetism for which $T_{\mu\nu} = F_{\mu \alpha} F_{\nu}{}^{\alpha}- \frac{1}{4} \eta_{\mu\nu} F_{\alpha \beta}F^{\alpha \beta}$ then the effective matter Lagrangian is
\ba
\label{MaxwellEft}
{\cal L}_{\rm matter}'&=&-\frac{1}{4} F_{\alpha \beta}F^{\alpha \beta} + \frac{2 C_{W^2}}{\mpl^4} \left(  {\rm Tr}(F^4) - \frac{1}{4} ({\rm Tr}[F^2])^2 \right) + {\cal O}(\mpl^{-4}) \,  ,\\
{\cal L}_{\rm matter}'&=&-\frac{1}{4} F_{\alpha \beta}F^{\alpha \beta} + \frac{C_{W^2}}{2\mpl^4} \left( F_{\alpha \beta}F^{\alpha \beta} \right)^2 +  \frac{C_{W^2}}{8\mpl^4} \left( \tilde F_{\alpha \beta}F^{\alpha \beta} \right)^2+ {\cal O}(\mpl^{-4}) \, .
\ea
Familiar arguments on the absence of superluminalities for photons in different backgrounds \cite{Adams:2006sv}, or equivalently positivity bounds applied {\bf assuming the graviton $t$--channel pole can be neglected} \cite{Bellazzini:2019xts} imply that the coefficients of both of the above dimension 6 operators are positive which is satisfied with the single condition
\be
C_{W^2} > 0 \, .
\ee

We emphasize that what has been performed here is an inverted logic as compared to what is typically considered  in the literature when imposing bounds on curvature operators. Rather than applying positivity bounds directly on the gravitational sector, we have applied it on scattering amplitudes of the matter sector. Subtleties related to the $t$--channel pole are of course equivalent in both cases, as it should be. The related arguments of \cite{Bellazzini:2019xts} similarly determine the positivity bounds on photon scattering.

\subsection{$TT$ amplitude}
\label{sec:TTamplitude}

The previous arguments allow us to impose constraints on the sign of the curvature--squared operators using positivity bounds applied on scattering amplitude of the matter sector provided the $t$--channel pole that appear in those amplitudes can be discarded.  A stronger form of these arguments follows from considering the $TT$--contributions from the matter Lagrangian to arise from the integration of massive spin--2 fields. Indeed, the effective matter Lagrangian may also be written as
\be
{\cal L}_{\rm matter}'={\cal L}_{\rm matter} + \frac{2 C_{W^2}}{\mpl^4}  \left( T_{\mu\nu} T^{\mu\nu} -\frac{1}{3} T^2 \right)+\frac{C_{R^2}}{\mpl^4} T^2+ {\cal O}(\mpl^{-4}) \, ,
\ee
which are the natural  combinations  following from the K\"all\'en--Lehman spectral representation for the $TT$ two--point function. This emphasizes the fact that this interaction could be viewed as the $TT$ amplitude obtained from integrating out massive spin--2 and higher states which naturally coupled to the stress energy through the $-1/3$ polarization factor and massive spin--0 states which could couple to the trace of the stress energy.
\subsubsection{Weakly Coupled UV Completion}

\label{Weaklycoupled}

To make the previous argument explicit, imagine a weakly coupled UV completion, with a potentially infinite tower of massive spin states. Let us imagine that matter couples to an effective metric build out of the Einstein frame zero mode metric $g_{\mu\nu}$ and some combination of all the other spin states. If the spin--states are only weakly excited, as would be expected in the regime of validity of the EFT, we may treat them as approximately linear, even while the zero mode metric $g_{\mu \nu}$ is nonlinear. Matter may then be taken to effectively coupled to $g_{\mu \nu}^{\rm eff} = g_{\mu \nu}+ \frac{1}{\mpl}\sum_i \alpha_i H^i_{\mu\nu}+\frac{1}{\mpl} \sum_j \beta_j \phi_j g_{\mu\nu}$ where $H^i_{\mu\nu}$ are any number of massive spin--2 particles of mass $M_i$ and $ \phi_j $ any number of scalar particles of mass $M_j$. Other spin states will not couple at this order. \\

The UV Lagrangian describing this set up may then be taken to be (ignoring Gauss--Bonnet terms)
\ba
&& {\cal L }_{\rm UV} \approx \sqrt{-g} \left[ \frac{\mpl^2}{2} R +  ( C^{\rm UV}_{R^2}- \frac{2}{3} C^{\rm UV}_{W^2} ) R^2 + 2 C^{\rm UV}_{W^2} R_{\mu\nu }^2  + \sum_i  \left[ \frac{1}{2} H_i^{\mu\nu} {\cal E} H^i_{\mu\nu}- \frac{1}{2} M_i^2 ({H^2_{i \, \mu\nu}}-H_i^2) \right]  \right.  \nn  \\
&& \left.  + \sum_j  \left[ \frac{1}{2} \phi_j \Box \phi_j- \frac{1}{2} M_j^2 \phi_j^2  \right] + {\cal L}_{\rm matter}  +\frac{1}{2\mpl}\sum_i \alpha_i H^i_{\mu\nu} T^{\mu\nu}+ \frac{1}{2\mpl} \sum_j \beta_j \phi_j g_{\mu\nu} T^{\mu\nu} + \dots \right]\,,
\ea
with ${\cal E}$ the covariant version of the Lichnerowitz operator and ellipses indicating higher order interactions for the additional spin fields. The obvious examples of this kind of effective Lagrangian are extra dimensional braneworld setups where matter is localized on a brane whose induced metric is not equivalent to the Einstein frame zero mode metric. The induced metric will indeed  include Kaluza--Klein modes $H_i^{\mu\nu} $ as well as potentially other scalar moduli fields $\phi_j$. \\

Now integrating out the massive fields to obtain the low energy effective theory, then due to the Fierz--Pauli structure of the mass term from the spin--2 fields we obtain the $-1/3$ factor, namely
\ba
 {\cal L }_{\rm IR} &\approx & \sqrt{-g} \Big[ \frac{\mpl^2}{2} R + ( C^{\rm UV}_{R^2}- \frac{2}{3} C^{\rm UV}_{W^2} ) R^2 + 2 C^{\rm UV}_{W^2} R_{\mu\nu }^2+{\cal L}_{\rm matter}  \\
 &+&\frac{1}{4\mpl^2 }\sum_i \frac{\alpha_i^2}{M_i^2} \( T^{\mu\nu} T_{\mu\nu}-\frac{1}{3} T^2 \) + \frac{1}{4\mpl^2} \sum_j \frac{\beta_j^2}{M_j^2} T^2 + \dots \Big]\,.\nn
\ea
Here we have made use of the fact that at leading order $\nabla^{\mu} T_{\mu\nu} \approx 0$ and so we the polarization tensors are simplified.
By rewriting the $T_{\mu\nu}^2$ and $T^2$ back in terms of curvature--squared interactions we may identify
\be
C^{\rm IR}_{W^2}=C^{\rm UV}_{W^2}+\sum_i \frac{\mpl^2}{8 M_i^2 } \alpha_i^2 \, ,  \quad C^{\rm IR}_{R^2}= C^{\rm UV}_{R^2}+ \sum_j \frac{\mpl^2}{4M_j^2 } \beta_j^2   \, .
\ee
that is
\be
\Delta C_{W^2} = \sum_i \frac{\mpl^2}{8M_i^2 } \alpha_i^2 >0 \, , \quad \Delta C_{R^2} = \sum_j \frac{\mpl^2}{4M_j^2 } \beta_j^2 >0 \, .
\ee
which gives the desired positivity properties. The equality could only be saturated if no fields coupled i.e. if $\alpha_i=\beta_j=0$.

\subsubsection{Generic UV Completion}

Although the previous argument was made explicitly for a weakly coupled UV completion, it follows equally well in general from the spectral representation for the $TT$ two--point function \cite{Dvali:2012zc}  for a conserved source, which would also apply when loops are included and hence for an arbitrary UV completion
\ba
\label{TT}
\Delta L_{TT} &=& \frac{1}{\mpl^4}\int_0^{\infty} \d \mu \ \rho_2(\mu) T^{\mu\nu} \frac{1}{\mu - \Box- i \epsilon} \( T_{\mu\nu} - \frac{1}{3} g_{\mu\nu} T\) \\
&+& \frac{1}{\mpl^4} \int_0^{\infty} \d \mu \ \rho_0(\mu) T \frac{1}{\mu - \Box- i \epsilon}  T  \, ,\nn
\ea
were standard unitarity arguments imply $\rho_2(\mu)  \ge 0$ and $\rho_0(\mu) \ge 0$. Crucially though this expression assumes that  no subtractions are necessary in writing this dispersion relation. The leading subtraction term can be directly absorbed into $C_{W^2}^{\rm UV}$ and $C_{R^2}^{\rm UV}$ so that we may formally write
\ba
&& C_{W^2}^{\rm IR} = C_{W^2}^{\rm UV} +  \frac{1}{2} \int_{0}^{\infty} \d \mu \frac{\rho_2(\mu)}{\mu} \, , \\
&& C_{R^2}^{\rm IR} = C_{R^2}^{\rm UV}+  \int_{0}^{\infty} \d \mu \frac{\rho_0(\mu)}{\mu} \, ,
\ea
with the understanding that $C_{W^2,R^2}^{\rm IR}$ are finite quantities. We see that $\alpha_i^2/M_i^2$ is replaced by the spin--2 spectral density divided by the spectral mass squared $\mu$ and $\beta_j^2/M_j^2$ replaced by the spin--0 spectral density divided by $\mu$. Again we see that
\be
\Delta C_{W^2} >0 \, , \quad \Delta C_{R^2}>0 \, .
\ee
We may define these coefficients at an arbitrary scale, which may be interpreted as the coefficients in the EFT defined with all states of energies greater than $M$ integrated out,  in the manner
\ba
&& C_{W^2}(M) = C_{W^2}^{\rm UV} +  \frac{1}{2} \int_{M^2}^{\infty} \d \mu \frac{\rho_2(\mu)}{\mu} \, , \\
&& C_{R^2}(M) = C_{R^2}^{\rm UV}+  \int_{M^2}^{\infty} \d \mu \frac{\rho_0(\mu)}{\mu} \, ,
\ea
so that the RG flow is finite (independent of subtraction/renormalization considerations) and positive in the sense
\be
C_{W^2}(M_1)-C_{W^2}(M_2) =\int_{M_1^2}^{M_2^2} \d \mu \frac{\rho_2(\mu)}{\mu} \, .
\ee
We thus see that standard spectral representation arguments imply the expectation that $\Delta C_{W^2}>0$ and related positivity bounds strengthen this to the expectation that $C_{W^2} > 0$. It is precisely with this sign that we found in (\ref{speedgravity1}, \ref{eq:speedwarped}) gravitational waves to be  superluminal on NEC preserving backgrounds unless one insists on having $C_{W^2}\equiv 0$ in which case Gravitational Waves would be luminal to this order (but not to higher orders as we see in Section~\ref{sec:Rcubed}).
\\

Related spectral representation arguments are given in \cite{Dvali:2012zc} and more recently in \cite{Mirbabayi:2019iae}, the latter chooses to neglect the $t$--channel pole by focusing on graviton pseudo--amplitudes which are essentially on--shell stress energy correlators \cite{Gillioz:2018kwh}. Similar arguments applied to the Gauss--Bonnet term or equivalent Weyl squared term in higher dimensions are given in \cite{Cheung:2016wjt} with the same implied choice of sign. S--matrix positivity arguments have been applied to quartic curvature interactions in \cite{Bellazzini:2015cra} complementing previous superluminality arguments \cite{Gruzinov:2006ie}. Entropic arguments that constrain the curvature--squared terms are given in \cite{Cheung:2018cwt} and are consistent with these implied signs.

\subsubsection{Neglecting the $t$--Channel Pole}

The K\"all\'en--Lehman spectral representation for a 2--tensor can also include contributions from massless graviton exchange. In the above we did not include this as we have intentionally written the action \eqref{decoupling} in a representation in which the massless graviton has not been integrated out. Had we done so then we would have obtained an effective matter Lagrangian
\be
{\cal L}_{\rm matter}''={\cal L}_{\rm matter}+ \Delta L_{TT}' \, ,
\ee
where
\ba
\label{DeltaTpp}
\Delta L_{TT}'  &=&\frac{1}{2 \mpl^2} T^{\mu\nu} \frac{1}{-\Box- i \epsilon}  \( T_{\mu\nu} - \frac{1}{2} g_{\mu\nu} T\) \\
&+& \frac{1}{\mpl^4}\int_0^{\infty} \d \mu \ T^{\mu\nu} \frac{\rho_2(\mu)}{\mu - \Box- i \epsilon} \( T_{\mu\nu} - \frac{1}{3} g_{\mu\nu} T\) + \frac{1}{\mpl^4} \int_0^{\infty} \d \mu \ T \frac{\rho_0(\mu)}{\mu - \Box- i \epsilon}  T \nn\,,
\ea
which is the full K\"all\'en--Lehman spectral representation between two conserved sources. The first term in \eqref{DeltaTpp} is of course the term that gives the pole terms in \eqref{phiphiamp}, in particular the massless $t$--channel pole which is responsible for the problems with applying standard forward limit dispersion relations. \\

In the above representation \eqref{DeltaTpp}, it is arguably obvious why we can ignore the contribution of the $t$ channel pole. Unitarity imposes positivity of $\rho_2(\mu) $ and $ \rho_0(\mu) $ through the requirement ${\rm Im}\( \Delta L_{TT}' \) \ge 0$ regardless of whether the massless pole part is present. This will allow us to determine a positivity bound for $C_{W^2}$ and $C_{R^2}$. This argument is slightly different from that given in \cite{Bellazzini:2019xts} since it focuses on only those interactions that be written in terms of a single $T_{\mu\nu}$, nevertheless it explains at least in part  why one would expect the result of \cite{Bellazzini:2019xts} with regards the neglect of the $t$--channel to be correct. The recent related discussion in  \cite{Mirbabayi:2019iae} similarly argues for $C_{W^2}>0$ at the level of scattering amplitudes, neglecting the contribution from the massless graviton by fiat by focusing on graviton pseudo--amplitudes.
\\

The real issue however is not the $t$--channel pole per se, but the required number of subtractions of the remaining terms. The dispersion relation \eqref{DeltaTpp} is only valid if the integrals
\be
\label{integrals}
\int_0^{\infty} \d \mu \frac{\rho_2(\mu)}{\mu} \, \quad \text{ and } \quad \int_0^{\infty} \d \mu \frac{\rho_0(\mu)}{\mu} \,
\ee
converge. If not then it is necessary to perform subtractions. For instance, performing one subtraction is equivalent to rewriting $\Delta L_{TT}' $ as
\ba
 \Delta L_{TT}'  &=&\frac{1}{2 \mpl^2} T^{\mu\nu} \frac{1}{-\Box- i \epsilon}  \( T_{\mu\nu} - \frac{1}{2} g_{\mu\nu} T\) + a_2 T^{\mu\nu} \( T_{\mu\nu} - \frac{1}{3} g_{\mu\nu} T\) + a_0 T^2   \\
&+& \frac{1}{\mpl^4}\int_0^{\infty} \d \mu \ T^{\mu\nu} \frac{\rho_2(\mu) \Box}{\mu (\mu - \Box- i \epsilon)} \( T_{\mu\nu} - \frac{1}{3} g_{\mu\nu} T\) + \frac{1}{\mpl^4} \int_0^{\infty} \d \mu \ T \frac{\rho_0(\mu) \Box }{\mu (\mu - \Box- i \epsilon)}  T \nn\,,
\ea
where $a_2$ and $a_0$ are the subtraction constants. After a field redefinition, the addition of these constants is equivalent to adding $R^2$ and $R_{\mu \nu}^2$ counterterms, which are known to be needed already in one--loop calculations to remove divergences, in other words, the inclusion of $C_{R^2}^{\rm UV} $ and $C_{W^2}^{\rm IR}$. In the relations $C_{W^2}^{\rm IR} = C_{W^2}^{\rm UV} +  \frac{1}{2} \int_{0}^{\infty} \d \mu \frac{\rho_2(\mu)}{\mu} $ etc., the RHS is a difference of two infinite quantities and it is hence not possible to conclude positivity of the LHS. Hence the ability to apply positivity bounds with the $t$--channel pole neglected comes down to whether the integrals \eqref{integrals} converge. This is not surprising since it is equivalent at the scattering amplitude level to requiring some Froissart--like bound on the $t$--channel pole subtracted amplitude. Alternatively this comes down to the question of how many subtractions are needed in specifying a dispersion relation for the graviton two--point function.

\subsubsection{Higher Derivative Corrections}

The positivity bounds implied by the $TT$ amplitude argument also apply to higher derivative terms. In particular if one subtraction is sufficient for convergence (as is known to be the case at one--loop level), then expanding the dispersion relation \eqref{TT} in powers of $\Box/\mu$ we infer at next order
\be
\Delta L_{TT} = \dots +  \frac{1}{\mpl^4}\int_0^{\infty} \d \mu \rho_2(\mu) T^{\mu\nu}  \frac{\Box}{\mu^2}\( T_{\mu\nu} - \frac{1}{3} g_{\mu\nu} T\) + \frac{1}{\mpl^4} \int_0^{\infty} \d \mu \rho_0(\mu) T  \frac{\Box}{\mu^2}  T  + \dots\, ,
\ee
which after a field redefinition is equivalent to a higher derivative curvature correction
\be
\Delta L_{TT} = \dots + \(  \int_0^{\infty} \d \mu \frac{\rho_2(\mu)} {\mu^2}  \) R_{\mu\nu} \Box R^{\mu\nu} +\left(  \int_0^{\infty} \d \mu  \frac{\rho_0(\mu) }{\mu^2}  -\frac{1}{3} \int_0^{\infty} \d \mu \frac{\rho_2(\mu)} {\mu^2}\right) R \Box R  + \dots\, .
\ee
As a nontrivial check on this, comparing with the explicit one--loop effective action from massive particles of spin 0,1/2,1 given in \eqref{eq:1LEFT} or Appendix~\ref{app1:1loop} for spin--0 then we have
\ba
&& \frac{1}{12(2\pi)^4}\sum_i  \frac{d_2^{(s_i)}}{M_i^2} =  \int_0^{\infty} \d \mu \frac{\rho_2(\mu)} {\mu^2} \, , \\
&& \frac{1}{12(2\pi)^4}\sum_i  \frac{d_1^{(s_i)}}{M_i^2} = \int_0^{\infty} \d \mu  \frac{\rho_0(\mu) }{\mu^2}  -\frac{1}{3} \int_0^{\infty} \d \mu \frac{\rho_2(\mu)} {\mu^2} \, .
\ea
The positivity of $\rho_2(\mu)$ and $\rho_0(\mu) $ at all scales implies that
\be
d_2^{(s_i) }> 0 \, , \quad d_1^{(s_i)} + \frac{1}{3} d_2^{(s_i)} >  0 \, .
\ee
From the results of Table~\ref{table:1-loopcoefs}, or from \eqref{eq:1loopR3} for spin--0, we see that all computed values of $d_2$ are positive and similarly $d_1^{(s_i)} + \frac{1}{3} d_2^{(s_i)} =(\frac{17}{840} , \frac{1}{840} , \frac{3}{280} )>0$ for $s_i = (0,1/2,1)$. These results are nontrivial since several of the $d_1^{(s_i)}$ are negative. Again we emphasize that since the one--loop effective action is finite at this order, \eqref{eq:1LEFT}   needs no renormalization counterterms which corresponds to the statement that the dispersion relation \eqref{TT} does not need any subtractions at this order.

\subsubsection{EFT matching}
In Section~\ref{sec:freq}, we have shown how already within the truncated EFT (only including the quadratic--curvature corrections), the sound speed is frequency dependent.  Moreover, to the level we have expanded the sound speed (while remaining in the regime $k\ll M$), the departures from luminality are always reduced at higher frequency regardless of the sign of $C_{W^2}$ and regardless of whether or not the background satisfies the NEC (i.e. regardless of whether the speed is sub or super luminal at low frequencies). Putting those arguments aside,
the clearest argument that the speed of sound returns to unity at high energy is obtained from the EFT matching. The speed we have calculated reflects the tree level speed identified from a Wilsonian effective action in which states heavier that some mass scale $M$ have been integrated out. As we have argued, in general we anticipate
\be
\Delta C_{W^2} = C_{W^2}(M_1) - C_{W^2}(M_2)= \int_{M_1^2}^{M_2^2} \d \mu \frac{\rho_2(\mu)}{\mu} >0 \, , \quad M_1 < M_2
\ee
where $C_{W^2}(M)$ denotes the associated coefficient in an EFT with masses above $M$ integrated out.  It follows that the associated speed of gravitational waves on FLRW defined in a given EFT is
\be
c_s^2(M) = 1+\frac{16 C_{W^2}(M) (-\dot H)}{\mpl^2} +\mathcal{O}\(\frac{H^4}{\mpl^4}\)
\ee
and so
for any fixed $\dot H$ (or $H'(y)$ in a static wrapped geometry), independently of the sign of $\dot H$ (resp. $H'(y)$) or the sign of $C_{W^2}$,
\be
|1-c_s^2(M_2)|  < |1-c_s^2(M_2)|\, , \quad  {\rm for }\quad M_2 > M_1\,.
\ee
In other words integrating back in the heavy modes reduces the departure of the speed of sound from unity. The need to recover Lorentz invariance strongly suggests \cite{deRham:2018red}
\be
\lim_{M \rightarrow \infty } C_{W^2}(M) = C_{W^2}^{\rm UV} =0 \, .
\ee

\subsection{Curvature Couplings to Light Fields}

\label{dim6lightfields}

In the previous discussion we assumed that the light fields that generate the cosmological background are minimally coupled to gravity. For instance, as a simple model the light field Lagrangian may be that describing a minimally coupled scalar field $\psi$ with a potential $V(\psi)$ or as a model of a perfect fluid a $P((\p \psi)^2,\psi)$ Lagrangian or it may be radiation from a Maxwell field, inflaton, quintessence etc.. To be concrete let us consider the example of the former so that we may take
\be
\L^{\rm (l.e.)}_\psi(g, \psi) = - \frac{1}{2} (\partial \psi)^2- V(\psi) \ .
\ee
It is of course consistent to imagine that these light fields also have non--minimal couplings to gravity. These may arise if the light field is itself nontrivially coupled to the heavy field, and then on integrating out the heavy fields we generate new curvature interactions for the light fields. Following the discussion in Section~\ref{positivitysec} we already know that interactions of the form $(\nabla \psi)^4$ could be viewed as having arise from field redefinitions. Far less trivial are the following dimension--6 operators which will contribute at the same order as the curvature--squared corrections,\footnote{A special case of these interactions has been considered in \cite{Hollowood:2016ryc} from the perspective of time-delays/advance. Our discussion here complements those.}
\be
\label{adddim6}
\Delta \L^{\rm (l.e.)}_\psi(g, \psi) = \frac{C_4}{\mpl^2} G^{\mu\nu} \nabla_{\mu} \psi \nabla_{\nu} \psi+  \frac{C_5}{\mpl^2} R (\nabla \psi)^2 + C_6 R \frac{U(\psi)}{\mpl^2}\, ,
\ee
where $U(\psi)$ is a function of $\psi$ with the same overall scale as $V(\psi)$. The addition of these operators modifies the background equations of motion at order $1/\mpl^2$ and these modifications needs to be accounted for in analyzing the perturbations. Following almost verbatim the previous recipe we find that the speed of propagation for tensors with the addition of these three non--minimal interactions becomes
\be
c_s^2=1+\frac{4( C_4 + 4C_{W^2}) (-\dot H)}{\mpl^2} +\mathcal{O}\(\frac{H^4}{\mpl^4}\)\,.
\ee
It would be tempting to suppose that the value of $C_4$ should be chosen in terms of $C_{W^2}$ so that either $c_s^2 \le 1$ or indeed by taking $C_4 = - 8 C_{W^2}$, $c_s^2=1$ so as to enforce GWs (sub)luminality. The problem with this is that it presupposes that the coefficient of whatever field is driving the cosmological expansion, is tied to the field content that has been integrated out. But a priori, the field content driving the expansion could be arbitrarily weakly coupled to the fields that have been integrated out. For instance $\psi$ may represent dark matter or dark energy degrees of freedom, that may lie in some dark sector arbitrarily weakly coupled to Standard Model fields. Furthermore the precise light fields determining the cosmological expansion are epoch dependent, inflaton, radiation, dark matter, dark energy, and there is no reason to suppose each of these distinct light fields should be non--minimally coupled in the precise manner needed to enforce $c_s^2=1$. More importantly, however just as in Section~\ref{positivitysec}, positivity bounds (to the extent where the $t$--channel pole may be ignored)  lead to the inevitable conclusion that $ C_4 + 4C_{W^2}>0$ and hence that  gravitational waves remain superluminal.   \\

To see how positivity bounds constrain the sign of $ C_4 + 4C_{W^2}$,  we note that the operators in \eqref{adddim6} can be removed with a field redefinition similarly as in Section~\ref{lightcone}
\be
g_{\mu \nu} \rightarrow g_{\mu\nu} - \frac{2}{\mpl^4} ( C_4 \nabla_{\mu} \psi \nabla_{\nu} \psi - C_5 (\nabla \psi)^2 g_{\mu\nu} - C_6 U(\psi)  g_{\mu\nu} )  \, ,
\ee
at the price of introducing additional matter interactions. Let us consider a spectator field $\chi$ which we take to be a free massless scalar living on the original metric. The above field redefinition will induce interactions between the scalar $\chi$ and the field $\psi$. Further removing the $R^2$ and Weyl squared terms via the field redefinition \eqref{fieldredef}, then to this order we obtain Einstein gravity minimally coupled to an effective matter Lagrangian
\ba
\label{twoscalareft}
 {\cal L}_{\rm matter}' &=&-\frac{1}{2} (\nabla \psi)^2-V(\psi) -\frac{1}{2} (\nabla \chi)^2 + \frac{1}{\mpl^4} \( 4 C_6 U(\psi) V(\psi) + 4 (4 C_{R^2}-\frac{2}{3} C_{W^2} )V(\psi)^2 \) \quad \\
&+&  \frac{1}{2\mpl^4} (\nabla \psi)^2 \Bigg[ 2 C_6 U(\psi) + 2 (8 C_{R^2}-\frac{4}{3} C_{W^2} -C_4+4C_5) V(\psi)   \nn  \\
&+& (2C_{R^2}+\frac{8}{3} C_{W^2} +C_4 + 2C_5) (\nabla \psi)^2 +(4 C_{R^2}-\frac{8}{3} C_{W^2} - C_4 + 2C_5) (\nabla \chi)^2\Bigg] \nn\\
&+& \frac{1}{\mpl^4} ( \nabla \chi)^2 \Bigg[ C_6 U(\psi) + (8C_{R^2}-\frac{4}{3} C_{W^2}) V(\psi) + (C_{R^2}+\frac{4}{3} C_{W^2}) (\nabla \chi)^2 \Bigg]  \nn\\
&+&\frac{(4 C_{W^2} +C_4)}{\mpl^4}( \nabla^{\mu} \chi \nabla_{\mu} \psi)^2 \nn\,.
\ea

\subsubsection*{Subluminal Spectator}

We may now take the double scaling limit $\mpl \rightarrow \infty$, keeping $C/\mpl^2$ and $C_4/\mpl^2$ finite, so that $\cal L_{\rm matter}'$ may be taken as a non--gravitational field theory living on Minkowski spacetime.  The effective equation for linearized fluctuations of the spectator field is
\be
\partial_{\mu} \( Z^{\mu\nu} \partial_{\nu} \chi \) =0 \, ,
\ee
with the effective metric
\ba
Z^{\mu\nu} &=& \eta^{\mu \nu} \( 1 -(4 C_{R^2}-\frac{8}{3} C_{W^2} - C_4 + 2C_5) \frac{(\partial \psi)^2}{\mpl^2} -\frac{2}{\mpl^4} \( C_6 U(\psi) + (8C_{R^2}-\frac{4}{3} C_{W^2}) V(\psi) \)  \) \nn \\
&-&\frac{2}{\mpl^4} (4 C_{W^2} +C_4) \partial^{\mu} \psi \partial^{\nu} \psi \, .
\ea
The effective speed of propagation on a background in which $\psi(t)$ is time--dependent is seen to be to this order
\be
c_s^2(\chi) = 1-\frac{2( C_4 + 4C_{W^2}) \dot \psi^2}{\mpl^4} +\mathcal{O}\(\mpl^{-4}\) =1-\frac{4( C_4 + 4C_{W^2}) (-\dot H)}{\mpl^2} +\mathcal{O}\(\mpl^{-4}\) \,.
\ee
Thus in performing the field redefinition to the frame in which gravity is minimally coupled, and hence gravitational waves are luminal, we have made the spectator field propagate subluminally by exactly the same amount. This is consistent with the general expectation that the ratio
\be
\frac{c_s^2(\rm tensors)}{c_s^2(\rm spectator)}
\ee
is field frame independent.
Hence demanding that in this decoupling limit that the spectator fields fluctuations are (sub)luminal requires
\be
C_4 + 4C_{W^2} \ge 0 \, .
\ee

\subsubsection*{Positivity Bounds}

We may provide a better argument by focusing on S--matrix positivity bounds applied to $\phi \psi \rightarrow \phi \psi $ scattering, assuming as per our previous discussion that we can neglect the massless $t$--channel pole. Regardless of the choice of potentials $V(\psi)$ and $U(\phi)$ we can see by power counting derivatives that the only terms in the Lagrangian \eqref{twoscalareft} that will potentially contribute to the twice subtracted scattering amplitude at tree level are (neglecting again the contribution from exchange of massless gravitons)
\be
 {\cal L}_{\rm matter}' \supset  \frac{(4 C_{W^2} +C_4)}{\mpl^4} ( \nabla^{\mu} \chi \nabla_{\mu} \psi)^2+\frac{(4 C_{R^2}-\frac{8}{3} C_{W^2} - C_4 + 2C_5)}{2\mpl^4} (\nabla \psi)^2 (\nabla \chi)^2
\ee
which gives a contribution of the form
\ba
 \A_{\psi \phi \rightarrow \psi \phi}(s,t) &\supset &
\frac{(4 C_{W^2} +C_4)}{\mpl^4} \(\frac{1}{2} (s-m_{\psi}^2)^2 +\frac{1}{2} (u-m_{\psi}^2)^2  \) \\
&+& \frac{(4 C_{R^2}-\frac{8}{3} C_{W^2} - C_4 + 2C_5)}{2\mpl^4}  t ( t -2 m_{\psi}^2) \, ,\nn
\ea
where $u= 2 m_{\psi}^2-t-s$ and $m_{\psi}$ is that mass of $\psi$.
Consequently
\be
\frac{1}{2} \frac{\p^2}{\p s^2} \A'(s=0,t=0) = \frac{(4 C_{W^2} +C_4)}{\mpl^4} \, ,
\ee
and so a standard application of the forward limit positivity bounds implies \cite{Pham:1985cr,Ananthanarayan:1994hf,Adams:2006sv}
\be
C_4 + 4C_{W^2} > 0 \, .
\ee
The equality cannot be saturated since the right hand side of the dispersion relation is determined from the total scattering cross section which is necessarily non--zero. Such considerations would then ensure that in this field frame the spectator field fluctuations are necessarily subluminal, or that in the original field frame the gravitational waves are necessarily superluminal. Although we have run this argument introducing a light spectator field, we could equally apply it to any Standard Model field. Indeed the role of $\chi$ could have been played by the Higgs field introducing a mass for $\chi$ will not change the result.

\section{Effects from Higher--Dimension Curvature Operators}

\label{sec:Rcubed}

In the previous section, we discussed the effect of the leading dimension--4 operators in the EFT expansion, the curvature--squared corrections. In the context of loop corrections from matter the precise coefficients of the curvature--squared terms cannot be determined since they arise logarithmically divergent and must be renormalized by introducing an appropriate counterterm $C_{R^2,W^2}^{\rm UV}$. The resulting renormalized coefficient $C_{R^2,W^2}^{\rm IR}$ can be consistently chosen to take any value without contradicting the requirements of consistency of the EFT. We have argued from positivity bounds that it is reasonable to suppose $C_{W^2}^{\rm IR} > 0$, which would then lead to superluminal gravitational waves on NEC preserving backgrounds. Nevertheless it remains technically possible that we have $C_{W^2}^{\rm IR} = 0$. Although a seemingly technically unnatural tuning, it would ensure that the non-luminal propagation we have uncovered so far is removed. This forces us to look to next order in the EFT expansion where things become more interesting. \\

In this section, we shall assume that the IR curvature--squared terms are set to zero, $C_{W^2}^{\rm IR}=C_{R^2}^{\rm IR}=0$ and that in such a basis, SM fields are minimally coupled to the metric. In this case any effect arising on the speed of propagation of gravitational waves will arise at next order in the EFT expansion, which in the present context means dimension--6 operators, i.e. curvature cubed and higher derivative curvature--squared terms, which we consider in subsection~\ref{sec:R3} before considering dimension--8 operators (fourth power of curvature) in subsection~\ref{sec:R4}. Unlike the curvature--squared terms, the curvature cubed terms that arise from matter loops are {\bf finite, calculable and have explicit dependence on the number of species}. There is no need to assume the existence of a UV contribution (at least in the absence of graviton loops) and so we can meaningfully consider unambiguous finite contributions to the sound speed. Moreover, unlike the case of the curvature--squared operators considered so far, there is no covariant and local field redefinition that can remove all the higher--order curvature operators that we consider in this section.

\subsection{Dimension--6 Curvature Operators}
\label{sec:R3}

\subsubsection{One--Loop Effective Action}
\label{sec:1lEFT}
The general form of the dimension--6 curvature operators that are expected to arise in a gravitational EFT are well known and can be parameterized by \cite{Metsaev:1986yb}
\ba
\label{eq:1LEFT}
\Gamma^{\rm (1-loop)}_{\rm dim-6}=\frac{1}{12(2\pi)^4}\sqrt{-g}&&\sum_i \frac{1}{M_i^2}\Bigg[
d_1^{(s_i)} R\Box R+ d_2^{(s_i)} R\mn \Box R^{\mu\nu}+d_3^{(s_i)}R^3+d_4^{(s_i)}R R\mn^2\\
&&+d_5^{(s_i)}R R_{\mu\nu\alpha\beta}^2+d_6^{(s_i)} R\mn^3+d_7^{(s_i)}  R^{\mu\nu}R^{\alpha\beta}R_{\mu\alpha\nu\beta}+d_8^{(s_i)} R^{\mu\nu}R_{\mu \alpha \beta\gamma}R_{\nu}{}^{ \alpha \beta\gamma}
\nn\\
&&+d_9^{(s_i)} R^{\mu\nu \alpha\beta}R_{\mu \nu \gamma \sigma}R_{\alpha \beta}{}^{ \gamma \sigma}+d_{10}^{(s_i)} R^{\mu}{}_\alpha{}^\nu {}_\beta R^\alpha{}_\gamma{}^\beta{}_\sigma R^\gamma{}_\mu{}^\sigma{}_\nu\nn
\Bigg]\,,
\ea
 \\
where $d_I^{(s_i)}$ denotes the contribution from integrating out a particle of mass $M_i$ and spin $s_i$. As usual, the above form can be simplified by using field redefinitions to remove for example all the $R$ and $R_{\mu\nu}$ terms (see \cite{Ruhdorfer:2019qmk} for a recent discussion), but doing so will only introduce interactions in the matter sector which capture the same basic one--loop process, such as those described in Fig.~\ref{Fig:loopTT}. Performing this field redefinition would take us out of the frame in which we have chosen to minimally couple SM fields and so we prefer to remain in this frame, being the natural one from the perspective of a path integral calculation.\\

Unlike the case for the curvature--squared corrections, other than the coefficients $d_1^{(s_i)}$ and $d_2^{(s_i)} $ there are no known positivity bounds that fix the signs of the remaining coefficients. That is because, even if we were to rewrite these interactions as pure matter ones, they would correspond to $T_{ab}T_{cd}T_{ef}$ interactions, and would only contribute (at tree--level) to $3-3$ scattering and higher order amplitudes, or to 3 point K\"all\'en--Lehman dispersion relations, for which clean statements of positivity are not known (although see \cite{Afkhami-Jeddi:2018own} for statements in the holographic/CFT context). \\

We can bypass this problem by however focusing on the explicit example of loop corrections from particles of spin $s_i \le 1$ for which the coefficients are known and are {\bf finite}. This finiteness is crucial since it tells us there is no need to add any counterterms at this order, and so their is no ambiguity about the signs of the resulting coefficients. For these dimension--6 operators, the explicit one--loop effective action was computed exactly in \cite{Avramidi:1986mj,Avramidi:1990je} for massive particles of spin $0,1/2$ and $1$, where the dimensionless coefficients $d_n^{(s_i)}$ are given in table~\ref{table:1-loopcoefs} of appendix~\ref{app1:1loop} and depend on the spin $s_i$ of the particle integrated. We reproduce these results for spin--0 explicitly in Appendix \ref{app1:1loop} using dimensional regularization for convenience. \\

As before, we shall see that the very existence of these dimension--6 operators leads to a sound speed for gravitational waves at low--energy which can differ from luminality. On a cosmological background they lead to corrections to the sound speed of order  $N \dot H H^2/\mpl^2 M^2$ where $N$ is the number of fields integrated out and $M$ their mass.  What is crucially different at this order is, depending on the field content of heavy modes, the speed of gravitational waves can effectively turn both {\it superluminal} and {\it subluminal}. This is true even for matter forced to respect the null energy condition, for signs we know to be consistent with positivity since they are derived from an explicitly unitary calculation from a local and well--behaved field content.

\subsubsection{Dimension--6 Curvature Operators on FLRW}

In what follows it will be convenient to define the {\rm effective} number $N^*_{s}$ of scalars $s=0$, vectors $s=1$ and spinors $s=1/2$ as
\ba
\label{eq:N*}
N^*_{s}=\sum_{\text{field $i$ of spin $s$}} \frac{M^2}{M_i^2}\,,
\ea
where we only include fields with masses above the scale of the EFT we are interested in, i.e. $M_i\gg H$ on FLRW.
Unless there is a large number particles $N\ggg 1$ at the same mass, we would typically expect this effective number to be dominated by the lightest massive particle beyond the low--energy EFT.  \\

Given our assumption that the IR curvature--squared terms have been set to zero, at energy scales below the mass $M$, then the leading terms in our gravitational EFT are
\ba
\label{eq:EFT1loop}
\L_{\rm IR}=\sqrt{-g}\left[\frac{\mpl^2}{2}R+\L^{\rm (l.e.)}_\psi(g, \psi)\right]+ \Gamma^{\rm (1-loop)}_{\rm dim-6} + \frac{1}{M^4}\mathcal{O}\((R^2, \nabla^2 R)^2\) \,,
\ea
where $\Gamma^{\rm (1-loop)}_{\rm dim-6}$ is given in \eqref{eq:1LEFT}, where the terms omitted are operators of dimension--8 and higher that are further considered in Section~\ref{sec:R4}. Expanding to quadratic order around an FLRW background, using the same conventions as in Section~\ref{Rsquared} with tensor modes normalized as in \eqref{eq:hij},  the contributions from the dimension--6 operators are of the form
\ba
\label{eq:Ldim6}
\L_{\rm dim-6}^{(hh)}=\frac{1}{12(2\pi)^4}\frac {a^2}{420 M^2}\ h \ \hat{O}_{\rm dim-6} \ h\,,
\ea
where the differential operator $\hat{O}$ includes up to fourth order in derivatives. More specifically, the operator can be put in the form
\ba
\label{eq:Odim61}
\hat{O}_{\rm dim-6}=\sum_{s} N^*_{s} \Bigg[
\frac 1{a^4}f_{1}^{(s)} \Box^2_\eta
+\frac 1{a^4}f_{2}^{(s)} \Box_\eta  \vec{\nabla}^2
+\frac 1{a^3}f_{3}^{(s)} \Box_\eta  \p_{\eta}
+\frac 1{a^2}f_{4}^{(s)} \Box_\eta\\
+\frac 1{a} f_{5}^{(s)} \p_{\eta}
+\frac 1{a^3} f_{6}^{(s)} \vec{\nabla}^2 \p_{\eta}
+\frac 1{a^2} f_{7}^{(s)} \vec{\nabla}^2
+ f_{8}^{(s)}
\Bigg]\nn\,,
\ea
where again $\eta$ is the conformal time and  $\Box_\eta=-\p_{\eta}^2+\vec \nabla^2$ is the d'Alembertian on Minkowski spacetime.
The functions $f_n^{(s)}$ depend on the background (and on the spin $s$ of the particles integrated). Their precise expressions are rather non--illuminating and are
provided in appendix~\ref{app2:FLRW}, (\ref{eq:f10}--\ref{eq:f8sp}). \\

As before, on pure de Sitter, the functions $f_{2,5,6,7}$ vanish and the one--loop contribution is a simple combination of flat d'Alembertian and effective mass terms acting on the tensor fluctuations, leading to the standard dispersion relation of the form $\omega^2=k^2 + m_0^2$. On FLRW however, the breaking of the maximal symmetry implies the existence of additional time--derivative operators that break the trivial relation between $\omega^2$ and $k^2$ in the dispersion relation.

\subsubsection{Modification of the Dispersion Relation}

Working perturbatively in the corrections from \eqref{eq:Ldim6}, we may use the equations of motion inferred from the standard Einstein and matter term $\L_{\rm EH, \psi}$  into \eqref{eq:Ldim6} so as to trade the  higher derivatives for lower ones as explained in Section~\ref{SpeedID}. This is performed explicitly in appendix~\ref{app2:FLRW} and we are then left with a {\it perturbative} second order action  of the form
\ba
\L^{(hh)}=\frac{\mpl^2}{4}a^2   h  \(\frac 1{a^2}\Box_\eta +2 H^2+ \dot H\) h + \frac{1}{12(2\pi)^4}\frac {a^2}{420 M^2}\ h \ \hat{O}_{\rm dim-6} \ h
\ea
with now
\ba
\label{eq:Odim62a}
\hat{O}_{\rm dim-6}=\sum_{s=0,1/2,1} N^*_{s} \Bigg[
\frac 1{a} \tilde f_{5}^{(s)} \p_{\eta}
+\frac 1{a^3} \tilde f_{6}^{(s)} \vec{\nabla}^2 \p_{\eta}
+\frac 1{a^2} \tilde f_{7}^{(s)} \vec{\nabla}^2
+ \tilde f_{8}^{(s)}
\Bigg]\,,
\ea
and the coefficients $\tilde f_{5-8}$ are given in (\ref{eq:ft5}--\ref{eq:ft8}). The friction term can then be removed as usual with a field redefinition, where the field redefinition is now momentum dependent (unless we are on de Sitter), of the symbolic form\footnote{In this symbolic notation, $H$ should be understood to also include all derivatives of $H$, so for instance $H^4$ is really a placeholder for $ H^4, H^2 \dot H, (\dot H)^2,  H \ddot H$ and $ \dddot H$.}
\ba
h \to  \left[1+ \frac{1}{\mpl^2 M^2}\(H^4 + \dot H \frac{k^2}{a^2}\)\right] h\,,
\ea
where the exact expression is provided in \eqref{eq:FR1} and \eqref{eq:FR2}. The apparent non--locality of this field redefinition does not cause a problem in identifying the speed of propagation, since it is perturbatively local and the associated Green's function should be determined perturbatively in it. The resulting equation of motion is then (symbolically) of the form
\ba
\label{eq:h3}
\(1+ \frac{A(k,\eta)}{\mpl^2 M^2}\)h'' + \(1+ \frac{B(k,\eta)}{\mpl^2 M^2}\)k^2 h + m_0^2 h =0 \,,
\ea
where the two functions $A(k,\eta)$ and $B(k,\eta)$ are symbolically of the form $A, B \sim H^4+\frac{k^2}{a^2}H^2$, but with precise coefficients that
differ away from de Sitter. We may thus infer an effective sound speed symbolically of the form $c_s^2=1+\frac{B-A}{\mpl^2 M^2}$, and derived explicitly in appendix~\ref{app2:FLRW} to be
\ba
\label{eq:cs2}
c_s^2=1-\frac{1}{12\times 105 (2\pi)^4  M^2 \mpl^2}&\Big(&
 2 (163 \Ns - 39 \Np - 659 \Nv)  H^2 \dot H\\
&-& (46 \Ns + 62 \Np + 530 \Nv) \dot H^2\nn\\
&+& (-93 \Ns + 46 \Np + 617 \Nv) H \ddot H\nn\\
&+& (-37 \Ns - 10 \Np + 57 \Nv) \dddot H
\Big)\nn\,,
\ea
and therefore departs from unity as soon as the background departs from pure de Sitter $H\ne $ const. At this level, we can also directly see that there is also no field content (no tuned values of $\Ns, \Np, \Nv$) that would lead to an exactly luminal speed for different cosmological epochs.

\subsubsection{Field Content Dependence}

\begin{figure}[h]
  \centering
  \includegraphics[width=\textwidth]{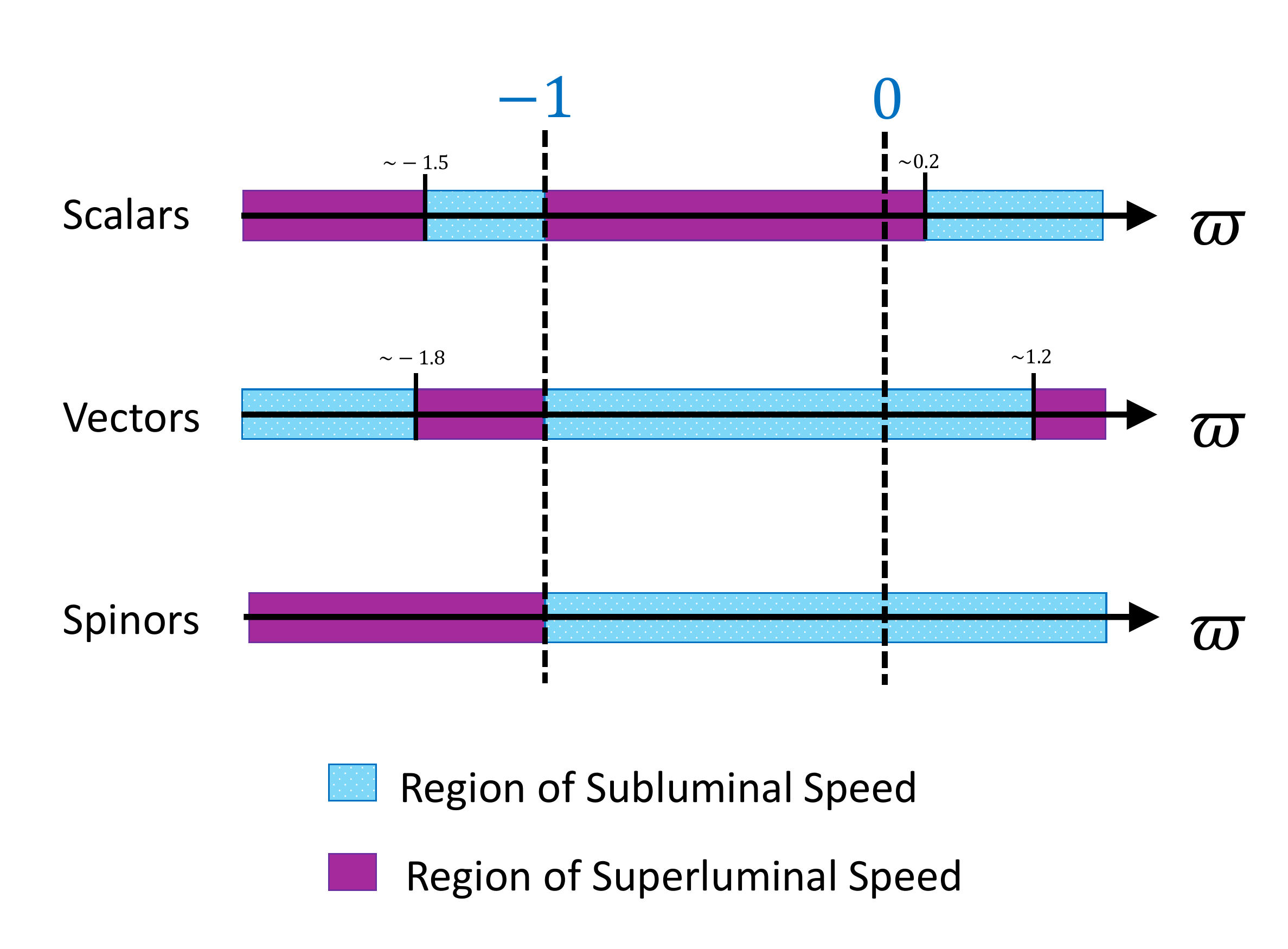}
  \caption{Regions of infinitesimal sub-- and super--luminal GW speed on FLRW with a constant equation of state parameter $\varpi$, from integrating out scalars, vectors or spinors.}\label{Fig:constw}
\end{figure}

Assuming a constant equation of state $\varpi$ on FLRW\footnote{In this section $\varpi$ represents the background equation of state parameter, not to be confused with the frequency $\omega$ of the GWs.},
the  effective speed of GWs is then
\ba
c_s^2&=&1-\frac{1}{12(2\pi)^4 140}\frac{H^4}{\mpl^2 M^4} (1+\varpi)\Bigg[
N_0^* (-349+1302\varpi+999 \varpi^2)\\
&&+6 N^*_{1/2}\(86+105 \varpi + 45 \varpi^2\)
+N_1^*\(3209-966 \varpi-1539 \varpi^2\)
\Bigg]\nn\,,
\ea
where $N_s^*$ is the effective number of spin--$s$ particles integrated out (as defined in \eqref{eq:N*}). The regions where the effective speed $c_s$ is sub-- vs super--luminal for scalars, vectors and spinors is depicted in Fig.~\ref{Fig:constw}.

\paragraph{Scalar vs Other Fields:}

Crucially, if the effective number of scalars integrated out dominates over that of vectors and spinors, then as we have defined it, $c_s$ would be {\it superluminal} for most of the late--time cosmological history of our Universe (since radiation--matter equality).
If on the other hand, the effect is dominated by vectors or spinors, then $c_s$ would be {\it subluminal} for the whole (standard) cosmological history of the Universe so long as one never crosses down to $\varpi<-1$ corresponding the breaking of the null energy condition. This is an entirely novel effect which did not arise at the previous order.

\paragraph{Subluminal Radiation Era:} Interestingly, independently on the precise field content, the finite contributions from integrating out heavier fields appears to lead to a {\it subluminal } speed for gravitational during the radiation era when $\varpi=1/3$.

\paragraph{Speed of GWs as a Discriminator:} In the past literature, an even small (unobservable) superluminal speed of gravitational waves has been commonly used as a discriminator between models or as a way to constrain parameters within an EFT (for instance related to the EFT of inflation or dark energy introduced in Refs.~\cite{Cheung:2007st,Gubitosi:2012hu}). In constructing the EFT for gravity, we are free to choose at what scale we wish to consider $M$ to be, provided we only consider backgrounds with $H\ll M$. With this in mind, we are free to integrate out Standard Model particles, including the electron and neutrino. In this family of EFTs, it will be that EFT defined with the lowest mass $M$ for which the dimension--6 operators that scale $1/M^2$ will have the largest effect. If we were to demand the gravitational waves to be subluminal, it would ``rule out" any model that postulates the existence of a scalar field with mass between the Hubble parameter and the neutrino mass, unless that particle was also accompanied with either vectors or spinors of comparable or lower mass. Applying this logic to current late--time cosmology with $H\sim 10^{-33}$eV, this would discriminate against any model that carries a scalar field with mass $m$ between say $10^{-30}$eV and $10^{-3}$eV, unless other fermions or vectors of comparably low mass were also included. For concreteness, in the absence of vectors, one should have $N^*_{0}\lesssim 0.2 N^*_{1/2}$ to avoid $c_s$ being superluminal as $\varpi$ approaches $-1$ (from above). For one spinor and one scalar field, this requires the mass of the scalar to be at least twice that of the spinor.

\paragraph{Implications for light Scalar Dark Matter:}
The search for dark matter has inspired the development of many light or even ultra--light scalar field models. Whether those scalar particles are charged or not, and whether they are pseudo--scalar is irrelevant for this discussion as all that matters here is the coupling to gravity. The considerations laid here could then potentially impact some of the following scalar models of dark matter such as axion or axion--like dark matter as well a fuzzy cold dark matter.
 In particular axion dark matter \cite{Holman:1982tb} (see \cite{Marsh:2017hbv} for a review) have opened a  quest for a multitude of experimental searches \cite{Graham:2015ouw}. The theoretical mass window for axion or axion--like dark matter spans over many orders of magnitude, but the open range is typically considered to be around $10^{-6}-10^{-2}$eV, while it is suggested that string theory may favor lower masses \cite{Graham:2015ouw}.
 As for fuzzy cold dark matter \cite{Hu:2000ke} their postulated mass could be as low as $10^{-22}$eV and within this logic would require equally low--mass spinors or vectors to avoid a small superluminal speed for GWs.

\paragraph{Speed of GWs as a Discriminator Redux:}
It is however clear from the discussions related to positivity bounds in previous sections for the curvature--squared terms (see Section~\ref{sec:Super}), that one should take great caution prior to jumping to any conclusion and using the presence, or absence, of superluminal speed for gravitational waves as a discriminator. Indeed, as we have seen in Section~\ref{sec:Super}, a superluminal speed for gravitational waves a low energy may not necessarily be in  conflict with a {\bf causal} UV completion and may sometimes even be favoured. Perhaps a  more appropriate criterion would be to require  that gravitational waves propagate faster than the lightcone to which matter minimally couples. This would then be the case for most of the recent cosmological history of our Universe if the curvature--squared corrections are included or if fields of spin--0 dominate the contributions to the curvature--cubed corrections. However applying such a criterion would also remain puzzling during the radiation era $\varpi=1/3$, where modes are always subluminal, unless one relies on the curvature--squared corrections.
It may be possible that it is not consistent in the given EFT to tune curvature--squared terms to zero, and that their positive contribution should dominate over the negative contribution from the dimension--6 curvature operator. For this to be true we need
\be
C_{W^2}\frac{|\dot H|}{\mpl^2} \gg N_* \frac{| \dot H|H^2}{M^2 \mpl^2}\,,
\ee
essentially at all scales $M$ as long as $H \ll M$. In other words
\be
C_{W^2} \gg N_* \frac{H^2}{M^2} \, .
\ee
Given the expectation that $C_{W^2} $ scales with the total number of species, and the EFT requirement that $H \ll M$, this condition is easily satisfied. Nevertheless from a low energy perspective there appeared to be nothing wrong with tuning $C_{W^2} =0$.

\subsection{Dimension--8 Curvature Operators}
\label{sec:R4}

Following from the previous logic, we may also inspect the effect of dimension--8 curvature operators (fourth power of curvature) on the graviton speed of sound. Such operators are known to arise in the effective action for string theory from tree level $\alpha'$ corrections \cite{Gross:1986iv,Giusto:2004xm}.
These were considered in \cite{Gruzinov:2006ie} for Ricci flat backgrounds with the following effective Lagrangian
\ba
\label{stringy}
\L=\sqrt{-g}\left[\frac{\mpl^2}{2}R+\frac{1}{M^4}\(c_1 (R_{\mu\nu\alpha\beta}^2)^2+c_2\, {\rm GB}^2\)\right]\,,
\ea
where ${\rm GB}$ is the Gauss--Bonnet term defined in \eqref{eq:GB}. Then following the same procedure as highlighted in Section~\ref{SpeedID} as applied for curvature--squared operators in Section~\ref{Rsquared} and for curvature--cubed operators in Section~\ref{sec:R3}, we find that on introducing a minimally coupled matter field (e.g. dilaton) which sources an FLRW (or equivalently on a space--dependent) background, we find a departure of the gravitational waves sound speed given by
\ba
c_s^2=1+\frac{384}{\mpl^2 M^4}\Bigg[
&c_1&  \(-8 \dot H H^4+ 3 \ddot H H^3+6\dot H^2 H^2+\dddot H H^2+5 H \dot H\ddot H + \dot H \dddot H+\ddot H^2\)\\
+&c_2&  \(-4 \dot H H^4+ 3 \ddot H H^3+10\dot H^2 H^2+\dddot H H^2+6 H \dot H\ddot H +\ddot H^2\)
\Bigg]\,.\nn
\ea
For constant equation of state parameter $\varpi$, this leads to a sound speed
\ba
c_s^2=1+\frac{144 H^6}{\mpl^2 M^4}\(1+\varpi\)
\(c_1 (19+15\varpi)(5+6\varpi+9\varpi^2)
-4c_2 (1+3\varpi)(17+15\varpi)\)\,.
\ea
It is clear that if the quadratic and cubic curvature operators were set to zero, then demanding the sound speed to be subluminal for NEC preserving backgrounds  (i.e. for $\varpi>-1$) would require $c_1\leq0$ which is in tension with the requirement found in \cite{Gruzinov:2006ie} for Ricci--flat backgrounds (unless we restricted ourselves to $c_1=c_2\equiv0$). Indeed in Ref.~\cite{Gruzinov:2006ie} it was found that subluminality on Ricci--flat backgrounds imposed $c_1,c_2\ge 0$. \\

If on the other hand one required that gravitational waves propagate faster than the lightcone to which matter minimally couples, this would require $c_1\ge 0$ and $-1.49 c_1\lesssim c_2\lesssim 1.09 c_1$.
Interestingly, it was argued in \cite{Gruzinov:2006ie} that considering the explicit quartic Riemann corrections from the string theory example proposed in \cite{Giusto:2004xm} and compactifing on a four--dimensional flat manifold would lead to $c_1=c_2>0$, which in the analysis provided on FLRW would  be compatible with a superluminal sound speed for gravitational waves. \\

A quite different result would arise if we first perform a field redefinition to remove any $R_{\mu\nu}$ and $R$ terms in \eqref{stringy} and then couple matter minimally to that new metric. In this case \eqref{stringy} reduces to a quartic Weyl operator which will not affect the speed of gravitational waves by virtue of the vanishing of the Weyl tensor on FLRW. This is an example of how the specific coupling to the light fields that source the background expansion is crucial in the analysis.

\section{Discussion}
\label{sec:discussion}

We have shown that in the standard effective field theory treatment of General Relativity coupled to matter, the low energy speed of gravitational waves, defined precisely in Section~\ref{SpeedID}, is inevitably different than unity on a background which spontaneously breaks Lorentz invariance, such as FLRW. The precise origin of this effect could be loop corrections from matter fields, or from higher spin particles $s \ge 2$ that may arise in a given UV completion. The former are effects that can never been turned off and arise in any UV completion. The latter are known to already arise in the low energy effective action for strings as discussed in Section~\ref{sec:R4}.
Perhaps surprisingly, for natural expectations of signs of Wilson (interaction) coefficients, the gravitational waves are generically found to be superluminal with respect to the metric with which the matter driving the background expansion is minimally coupled.  This effect is in some sense a pure gravitational analogue of the well known property of the low energy EFT for QED on a curved spacetime \cite{Drummond:1979pp,Lafrance:1994in}. \\

As is well known, low energy superluminal group or phase velocity are themselves not a direct sign of acausality. The front velocity, corresponding to the large frequency limit of the phase velocity is the speed at which information travels and a superluminal group velocity does not imply superluminal propagation of information. A discussion of this point is presented in \cite{Milonni,Brillouin,deRham:2014zqa} and has been emphasized in discussions of QED in curved spacetime \cite{Hollowood:2007kt,Hollowood:2007ku,Hollowood:2008kq,Hollowood:2009qz,Hollowood:2010bd,Hollowood:2010xh,Hollowood:2011yh,Hollowood:2012as}. In the present context, the ambiguity of metric field redefinitions means that there is no absolute definition of low energy speeds, only relative ones. The ratio of the speed of gravitational waves to an individual species of matter is invariant under field redefinitions. \\

Causality is likely better implemented by demanding S-matrix analyticity. At low energies, this criterion can be used to derive various positivity bounds that fix the signs of coefficients in the effective Lagrangian  \cite{Pham:1985cr,Ananthanarayan:1994hf,Adams:2006sv,deRham:2017avq,deRham:2017zjm}. If we assume that these bounds may be applied in the context of gravitational systems --- assuming following \cite{Bellazzini:2019xts} that it is possible to neglect the massless $t$-channel graviton pole --- then we have derived precise positivity bounds in Section~\ref{sec:Super} that enforce that regardless of field redefinitions, gravitational waves travel faster than allowed by the metric to which matter minimally couples. This apparent acausality is in fact seen after a field redefinition to be equivalent to the more prosaic requirement that matter fluctuations are (sub)luminal, and is thus usually regarded as a requirement of causality. \\

In our analysis we have focussed mainly on curvature corrections in the EFT, keeping the matter that sources the background relatively minimal. A more complete analysis could for example focus on the full effective theory of the Standard Model coupled to GR, (see \cite{Ruhdorfer:2019qmk} for an explicit discussion), but the examples considered in Section~\ref{dim6lightfields} support the universality of the connection between positivity bounds and superluminal gravitational wave propagation. Furthermore, in the case of low energy QED (or more generally the EFT of a $U(1)$ gauge field coupled to gravity), for which the general leading order EFT treatment is well known, positivity bounds have been applied recently in \cite{Bellazzini:2019xts},
and the signs are consistent with expectations when applied in FLRW. \\

Assuming no fine tuning of the EFT (no special properties of the UV completion), the magnitude of the departure of the propagation speed from unity is of order
\be
c_s^2 -1 \sim \frac{\dot H}{\Lambda^2}\,,
\ee
where $\Lambda$ is the EFT cutoff, i.e. either the mass scale of any heavy state integrated out, or the strong coupling scale of the theory. On de Sitter $\dot H=0$ and so $c_s=1$ as required by de Sitter invariance. Hence during inflation this effect is further slow roll suppressed. We have found that quite generally the magnitude and sign of the effect is controlled by the coefficient of the Weyl squared term in the effective Lagrangian. Given this, it is worth looking for independent arguments that may be used to constrain its sign and magnitude (see for example \cite{Cheung:2016wjt,Afkhami-Jeddi:2018own}). \\

If the leading curvature--squared terms in the EFT expansion are tuned to zero, as may be implied by a specific UV completion, then the dominant effect will come from higher order, whether dimension--6 or dimension--8 curvature operators. When this is the case, not only does the sign of the effect switches at the offset of the NEC but also becomes epoch dependent, a function of the precise equation of state of the Universe. The attempt to demand either universal sub--luminality or super--luminality of gravitational waves would in turn place strong constraints on the particle content in the Universe. \\

Our results have clear implications for cosmological effective field theory model building where it is common to assume that all fields (including tensor modes) are (sub)luminal in constraining the form of the effective action. Generically such a criterion is not well founded, and indeed it is not even invariant under field redefinitions. At best it can be implemented in some field frame, but as we have seen this is not the natural one to which we expect Standard Model fields to couple. Already in the case of the low energy effective theory for QED, it is known that backgrounds can be found in which different photon polarizations travel both superluminally and subluminally, undermining the existence of a single preferred field frame. What is needed is a better understanding of how causality, likely through S-matrix analyticity, could be used to constrain cosmological EFTs. There have been some attempts in the recent literature \cite{Baumann:2015nta,Afkhami-Jeddi:2018own,Melville:2019wyy,Baumann:2019ghk} although a clear understanding is absent due to the particular challenge of dealing with a massless graviton and the unclear meaning of analyticity on a curved spacetime. From our analysis it is clear that a significant role is played by the fully pole subtracted, elastic scattering amplitudes for matter fields. This is not so surprising given the similar role in non gravitational theories  \cite{Pham:1985cr,Ananthanarayan:1994hf,Adams:2006sv,deRham:2017avq,deRham:2017zjm}, nevertheless the gravitational extension has yet to be fully understood.

\bigskip
\noindent{\textbf{Acknowledgments:}}
We would like to thank John Donoghue for extremely useful discussions and comments.
AJT and CdR would like thank the Perimeter Institute for Theoretical Physics for its hospitality during part of this work and for support from the Simons Emmy Noether program. The work of AJT and CdR is supported by an STFC grant ST/P000762/1. CdR thanks the Royal Society for support at ICL through a Wolfson Research Merit Award. CdR is supported by the European Union's Horizon 2020 Research Council grant 724659 MassiveCosmo ERC--2016--COG and by a Simons Foundation award ID 555326 under the Simons Foundation's Origins of the Universe initiative, `\textit{Cosmology Beyond Einstein's Theory}'. AJT thanks the Royal Society for support at ICL through a Wolfson Research Merit Award. AJT would like to express a special thanks to the Mainz Institute for Theoretical Physics (MITP) off the DFG Cluster of Excellence PRISMA$^+$ (Project ID 39083149), for its hospitality and support.

\newpage

\appendix

\section{One--Loop Effective Action}
\label{app1:1loop}

\begin{figure}[h]
  \centering
  \includegraphics[width=\textwidth]{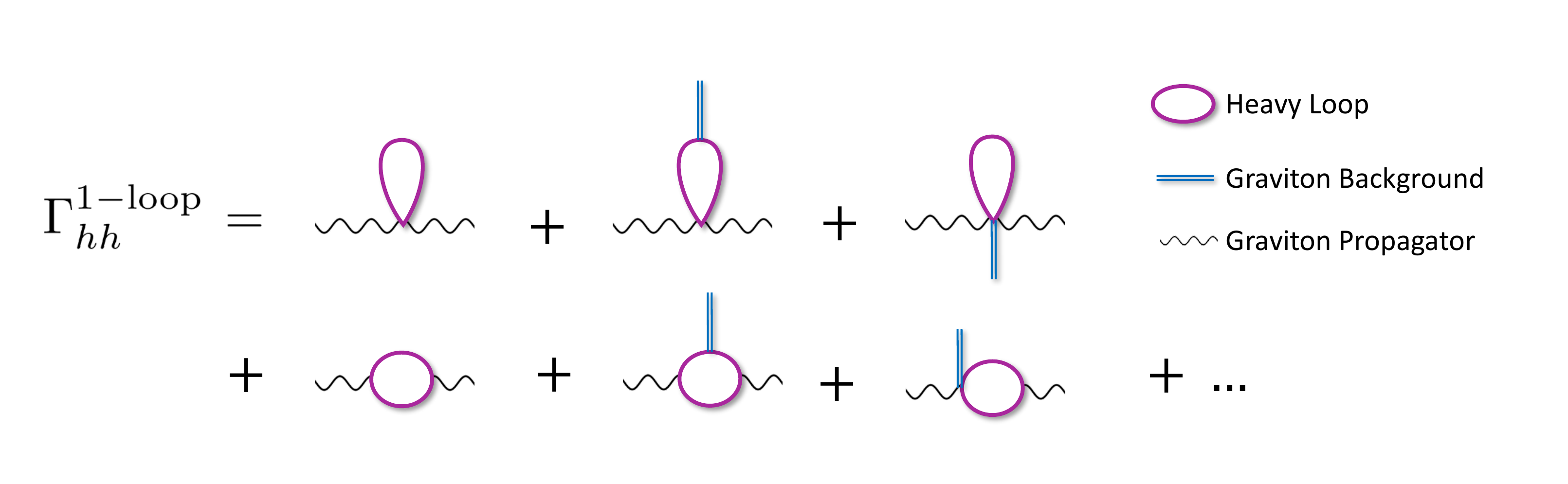}
  \caption{Example diagrams indicating how loops of heavy fields correct the propagation of gravitational waves in a background geometry. For weak backgrounds, the effect can be entirely accounted for by perturbative QFT in Minkowski spacetime (as we show below), giving identical results to covariant approaches.}\label{Fig:1LoopEFT}
\end{figure}

Here we show how to recover the one--loop effective action derived in \cite{Avramidi:1986mj} following a perturbative diagrammatic approach in the case of scalar fields, see Fig.~\ref{Fig:1LoopEFT}. The results are in complete agreements with those provided in \cite{Avramidi:1986mj}.

As we shall see below, by integrating (perturbatively) a massive scalar field minimally coupled to gravity, we recover the one--loop effective action \eqref{eq:1LEFT} with the precise same coefficients for the case of scalar fields. In Ref.~\cite{Avramidi:1986mj} the one--loop effective action for the spin--1/2 and --1 fields were also derived, and depending on the spin of the particle integrated out, the coefficients $c_n$ are given in  table~\ref{table:1-loopcoefs} below.

\begin{table}[h]
  $\phantom{.}$\hspace{-1.1cm}\begin{tabular}{|c|c c c  c  c  c  c  c  c  c |}
  \hline
  spin $s$ & $d_1$ & $d_2$ & $d_3$ & $d_4$ & $d_5$ & $d_6$ & $d_7$ & $d_8$ & $d_9$ & $d_{10}$ \\[3pt] \hline
  $0$ & $ 1/56 $ & $ 1/140 $ & $ 1/6^3 $ & $ -1/180 $ & $ 1/180 $ & $ -8/945 $ & $ 2/315 $ & $ 1/1260 $ & $ 17/7560 $ & $ -1/270 $ \\
  $1/2$ & $ -3/280 $ & $  1/28 $ & $ 1/864 $ & $ -1/180 $ & $ -7/1440 $ & $ -25/756 $ & $ 47/1260 $ & $ 19/1260 $ & $ 29/7560 $ & $  -1/108 $ \\
  $1$ & $ -27/280 $ & $  9/28 $ & $ -5/72 $ & $ 31/60 $ & $ -1/10 $ & $ -52/63 $ & $ -19/105 $ & $ 61/140 $ & $ -67/2520 $ & $ 1/18 $\\
  \hline
\end{tabular}
  \caption{Coefficients entering the dimension--6 operators in the one--loop effective action, for scalars, spinors and vectors. From \cite{Avramidi:1986mj}.}
  \label{table:1-loopcoefs}
\end{table}

\subsection{Minimally Coupled Scalar Field}

For concreteness, we shall integrate a scalar field $\varphi$ of mass $M$ minimally coupled to gravity,
\ba
\L[{\varphi}]=\sqrt{-g}\(-\frac 12 g^{\mu\nu} \p_\mu \varphi \p_\nu \varphi-\frac12 M^2 \varphi^2\)\,.
\ea
We work perturbatively in the metric perturbations about flat spacetime defined as
\ba
g\mn=(\eta\mn+\varsigma h\mn)^2=\eta\mn+2 \varsigma h\mn+\varsigma^2 h_{\mu\alpha} h_{\nu\beta}\eta^{\alpha\beta}\,,
\ea
where the parameter $\varsigma$ has been introduced for bookkeeping and
we do not yet commit to $h\mn$ being the tensor fluctuation as in what follows it will also carry the role of the background field.
To third order in perturbations for $h\mn$ (i.e. third order in $\varsigma$), one has
\ba
\sqrt{-g}&=&\frac 1{4!}\varepsilon^{abcd}\varepsilon^{a'b'c'd'}(\eta_{aa'}+\varsigma h_{aa'})
(\eta_{bb'}+\varsigma h_{bb'})
(\eta_{cc'}+\varsigma h_{cc'})
(\eta_{dd'}+\varsigma h_{dd'})\\
&=&1+\varsigma h+\frac {\varsigma^2}{2}([h]^2-[h^2])+\frac{\varsigma^3}{3!}([h]^3-3 [h][h^2]+2 [h^3])+\mathcal{O}(\varsigma^4)\\
g^{\mu\nu}&=&\eta^{\mu\nu}-2\varsigma h^{\mu\nu}+3 \varsigma^2 h^2{}^{\mu\nu}-4 \varsigma^3 h^3{}^{\mu\nu}+\mathcal{O}(\e^4)\,,
\ea
where square brackets represent the trace (wrt Minkowski) of a tensor.

The relevant graviton--scalar vertices are therefore
\ba
\Diagram{
& & \vertexlabel^{p} \\
& fuA \\
\vertexlabel_{h\mn} h &   \\
&  fdA\\
& & \vertexlabel_{q}
}
\quad &= &\frac 12 (M^2-p\cdot q) [h]\ p\cdot q+ h\mn p^\mu q^\mu\\[0.7cm]
\label{V2}
\Diagram{
\vertexlabel^{h\mn}& & & \vertexlabel^{p} \\
& hd & fuA \\
& hu &  fdA\\
\vertexlabel_{h\ab}& & & \vertexlabel_{q}
}
\quad &= &\frac 14 \((M^2-p\cdot q)([h]^2-[h^2])-6 h^2\mn p^\mu q^\mu+4 h h\mn p^\mu q^\mu \)\\[0.7cm]
\feyn{\vertexlabel^{h_{ab}}h} \hspace{-0.5cm}
\Diagram{
\vertexlabel^{h\mn}& & & \vertexlabel^{p} \\
& hd & fuA \\
& hu &  fdA\\
\vertexlabel_{h\ab}& & & \vertexlabel_{q}
}
\quad &= & \frac{1}{12}\Bigg(
(M^2-p\cdot q)([h]^2-[h^2])([h]^2-3[h][h^2]+2[h^3])\\
&+&6 ([h]^2-[h^2]) h\mn p^\mu q^\mu-18 h h^2\mn p^\mu q^\mu+24h^3\mn p^\mu q^\mu
\Bigg)\nn\,.
\ea

With those vertices in mind we can directly compute the relevant 1--loop scalar field contributions to the two and three point graviton. We do so using dimensional--regularization and summarize our conventions in what follows.

\subsection{Dimensional Regularization}

We work in $d=4-\e$ dimensions in what follows. Defining the following integrals $I^\ell_n$ as
\ba
I^\ell_n=\frac{\mu^{\e}}{M^{4+2(\ell-n)}}\int \frac{\d^d k}{(2\pi)^{d/2}}\frac{(k^2)^\ell}{\(k^2+M^2\)^n}\,,
\ea
we have in dimension--regularization,
\ba
&& I_0^\ell=0 \quad \forall\ \ell\\
&& I_1^0\equiv \bar I\equiv\frac{1}{\e}I_i+I_f\\
&& I_2^0=-I^0_1-2 I^0_3\\
&& I_3^0=-\frac{1}{4}\bar I_1\\
&& I_n^0=\frac{2(n-3)!}{(n-1)!}I_3^0\quad n\ge 4\\
&& I_n^\ell=I_{n-1}^{\ell-1}-I_n^{\ell-1}\,,
\ea
where we emphasize that both $I_1$ and $I_i$ are finite but $\bar I$ is not in the limit $\e \to 0$.

We will also make use of the following relations
\ba
&&\frac{\mu^{\e}}{M^{4+2(\ell-n)}}\int \frac{\d^d k}{(2\pi)^{d/2}}\frac{k_\mu k_\nu (k^2)^{\ell-1}}{\(k^2+M^2\)^n}=\frac 1 d \eta\mn I_n^\ell\\
&&\frac{\mu^{\e}}{M^{4+2(\ell-n)}}\int \frac{\d^d k}{(2\pi)^{d/2}}\frac{k_\mu k_\nu k_\alpha k_\beta (k^2)^{\ell-2}}{\(k^2+M^2\)^n}=\frac 1 {d(d+2)}
\( \eta\mn\eta\ab+\eta_{\mu \alpha}\eta_{\nu \beta}+\eta_{\nu \alpha}\eta_{\mu \beta}\) I_n^\ell\,,
\ea
and similarly for higher order integrals.

\subsection{Perturbative 1--Loop Contributions}

\subsubsection{Graviton One--Point Function}

For consistency, we start with the one--loop massive scalar field contribution to the one--point function, which is the leading term of the cosmological constant (and of course at the origin of the cosmological constant problem). The related Feynman diagram and  amplitudes are
\ba
\mathcal{A}_1 = \vcenter{\hbox{
\includegraphics[width=2cm]{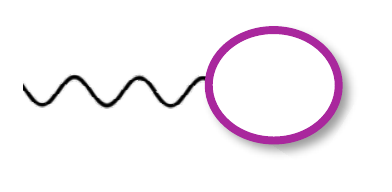}}}
=\frac{M^4}{4}\(\bar I +\frac{I_i}{4\e}\) h = \frac{M^4}{4}\(\bar I +\frac{I_i}{4\e}\) \delta_1 \sqrt{-g}\,,
\ea
this is a diverging contribution and leads to a contribution to the effective cosmological constant that scales as $M^4$. The order of magnitude of this cosmological constant is of course well--above that of the classical value we have considered so far. Tackling the cosmological constant problem is well beyond the scope of this work and for now we shall simply put this problem aside.

\subsubsection{Graviton Two--Point Function}

Two types of diagrams contribute to the two--point function. The first one, referred below as $\mathcal{A}_2^{(a)}$ and coming from the vertex \eqref{V2} is simply
\be
\mathcal{A}_2^{(a)}=\includegraphics[width=1.5cm]{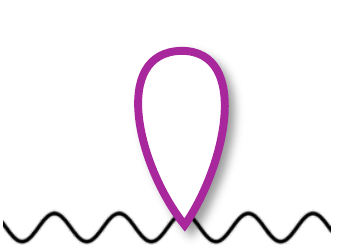}
=\frac{M^4}{4}\(\bar I +\frac{I_i}{4\e}\) \([h]^2-\frac{3}{2}[h^2]\)\,.
\ee
As for the second one, referred to as  $\mathcal{A}_2^{(b)}$, it involves a product of Feynman propagators, which, for the context of this work we simply deal by expansion in power of derivatives $p$ of the external legs,
\ba
\frac{1}{(k^2+M^2)((k-p)^2+M^2)}=\frac{1}{(k^2+M^2)^2}+2\frac{2 k\cdot p}{(k^2+M^2)^3}+\(\frac{4 (k\cdot p)^2}{(k^2+M^2)^4}-\frac{p^2}{(k^2+M^2)^3}\)+\cdots\,,\nn
\ea
and work up to sixth order in the external leg derivative.

Then as a perturbative expansion, the contribution of the second diagrams to the graviton two--point function is
\ba
\mathcal{A}_2^{(b)}&=&
\vcenter{\hbox{\includegraphics[width=2.5cm]{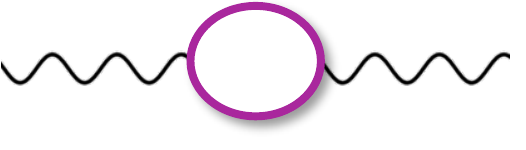}}}\\
&-&\frac{M^4}{8}\(\bar I +\frac{I_i}{4\e}\) \([h]^2-2[h^2]\)\nn \\
&+&\frac{M^2}{12}\bar I h^{\mu\nu}\left[
\Box h\mn-2 \p_\mu\p_\alpha h^\alpha_\nu+\p_\mu \p_\nu h -\eta\mn (\Box h-\p_\alpha \p_\beta h^{\alpha \beta})
\right]\nn\\
&+&\frac{1}{240}\(\bar I-I_i\)h^{\mu\nu}\left[
4\p_\mu \p_\nu \p_\alpha \p_\beta h^{\alpha \beta}-3 \Box \p_\mu \p_\nu h-2 \Box \p_\alpha \p_\mu h^\alpha_\nu+\Box^2 h\mn+3\eta\mn (\Box^2 h-\Box \p_\alpha \p_\beta h^{\alpha \beta})
\right]\nn\\
&-&\frac{I_1}{3360M^2}h^{\mu\nu}\Box\left[
12\p_\mu \p_\nu \p_\alpha \p_\beta h^{\alpha \beta}-11 \Box \p_\mu \p_\nu h-2 \Box \p_\alpha \p_\mu h^\alpha_\nu+\Box^2 h\mn+11\eta\mn (\Box^2 h-\Box \p_\alpha \p_\beta h^{\alpha \beta})
\right]\nn\\
&&+h\ \mathcal{O}\( \frac{\p^8}{M^4}\)\, h \nn\,,
\ea
where we immediately recognise the \Lic operator on the second line, corresponding to the leading expansion of the Einstein--Hilbert term $\sqrt{-g}R$.\\

Adding the one and two--point functions, we see that to the order we are working this precisely equivalent to the following effective action
\ba
\mathcal{A}_1+\mathcal{A}_2&\equiv& \frac{M^4}{4}\(\bar I +\frac{I_i}{4\e}\) \sqrt{-g}
+\frac{M^2}{6}\bar I \sqrt{-g}R+\frac{1}{240}\(\bar I-I_i\)\sqrt{-g}\(\frac 12 R^2+R\mn^2\)\\
&+&\frac{I_i}{6720M^2}\sqrt{-g}\(5 (\p R)^2+2 (D_\mu R\ab)^2\)+\mathcal{O}\(\varsigma^3, \frac{\p^8}{M^4} \)\nn\,,
\ea
where so far the right hand side should be understood as being only up to second order in the metric fluctuation. We see that all the terms on the first line diverge and can in principle be fully removed by appropriate renormalization procedure while the term on the second line is entirely finite.

\subsubsection{Graviton Three--Point Function}

We can follow the same procedure for the three point function. Three types of diagrams contribute at that level. The first one is a tadpole type and contains no derivatives acting on the external legs hence solely leading to a contribution towards the Cosmological Constant at cubic order in $\varsigma$
\ba
\mathcal{A}_3^{(a)}=
\vcenter{\hbox{\includegraphics[width=2cm]{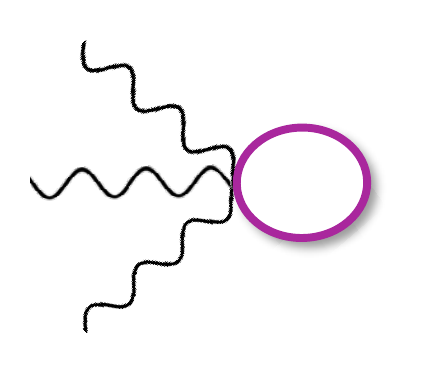}}}
=
\frac{M^4}{8}\(\bar I +\frac{I_i}{4\e}\) \([h]^3-4[h][h^2]+4[h^3]\)\,.
\ea
The next diagram, involving both a $hh \varphi \varphi$ and a $h \varphi \varphi$ vertex leads to contributions that are again computed performing a derivative expansion. Below we only present explicitly the leading order contributions  but they have been explicitly computed up to sixth order in derivative,
\ba
\mathcal{A}_3^{(b)}&=&
\vcenter{\hbox{\includegraphics[width=2.5cm]{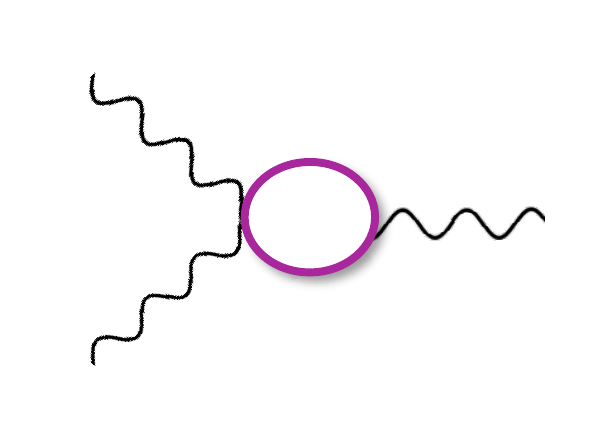}}}
 \nn \\
&=&
-\frac{M^4}{8}\(\bar I +\frac{I_i}{4\e}\) \([h]^3-5[h][h^2]+6[h^3]\) \\
\nn &&+ M^2 \bar I h^2 \p^2 h + (\bar I-\frac 12 I_i)h^2 \p^4 h+ \frac{I_i}{M^2}h^2 \p^6 h+\mathcal{O}\(h^2 \frac{\p^8}{M^4}h\)\,,
\ea
where the first line is exact while the second line is of course only symbolic but the accounts for the correct split between finite and divergent pieces.

Finally the diagram involving three $h \varphi \varphi$ vertices is the most challenging to account for and once again we only give its symbolic form in what follows (apart from the first line which is exact) even though it has been computed explicitly to sixth order in derivatives,
\ba
\mathcal{A}_3^{(c)}&=&
\vcenter{\hbox{\includegraphics[width=2.5cm]{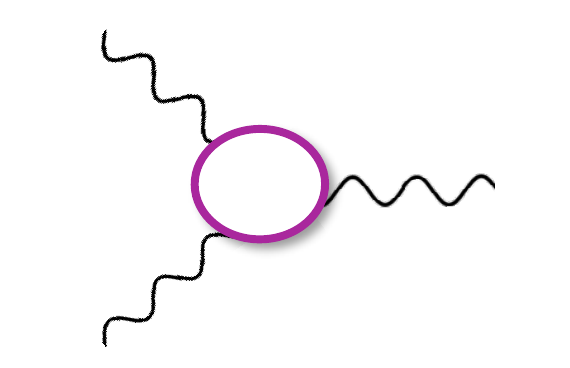}}}\nn \\
&=&\frac{M^4}{12}\(\bar I +\frac{I_i}{4\e}\) \(\frac 12 [h]^3-3[h][h^2]+4[h^3]\) + M^2 \bar I  h \p^2 h^2
+ (\bar I-\frac 12 I_i) h \p^4 h^2\\
&+& \frac{I_i}{M^2} h \p^6 h^2
+ \mathcal{O}\(h \frac{\p^8}{M^4}h^2\)\,.\nn
\ea
Combining the one, two and three point functions explicitly, one can check that up to cubic order in the metric perturbation and up to sixth order in derivatives, we obtain the following effective action (up to integrations by parts)
\ba
\label{eq:1loopR3}
\mathcal{A}_1+\mathcal{A}_2+\mathcal{A}_3&\equiv& \frac{M^4}{4}\(\bar I +\frac{I_i}{4\e}\) \sqrt{-g}
+\frac{M^2}{6}\bar I \sqrt{-g}R+\frac{1}{240}\(\bar I-I_i\)\sqrt{-g}\(\frac 12 R^2+R\mn^2\)\\
&+&\frac{I_i}{6720M^2}\sqrt{-g}\(5 (\p R)^2+2 (D_\mu R\ab)^2\)\nn\\
&-&\frac{I_i}{241920 M^2}\sqrt{-g}\Big(
47 R^3-80 R\mn^3-60 R R\mn^2+72 R^{\mu\nu}R^{ab}R_{\mu a \nu b}\nn \\
&&+56R_{\alpha \mu \beta \nu}R^{\alpha a \beta b}R_{ab}{}^{\mu\nu}
-40R_{\alpha \mu \beta \nu}R^{\alpha a \beta b}R_{a}{}^\mu{}_{b}{}^{\nu}
+114 R R_{abcd}^2
\Big)
+\mathcal{O}\(\varsigma^4, \frac{\p^8}{M^4} \)\nn\,.
\ea
Where again, only the terms on the first two lines include running and divergent pieces and could be entirely removed via appropriate renormalization procedure  while the remainder is finite. In particular $\bar I$ contains running and divergent pieces, $I_i/\e$ is divergent but $I_i$ itself is finite and does not run. One can check that they precisely match the coefficients given in Table~\ref{table:1-loopcoefs} for scalars as derived by \cite{Avramidi:1986mj}, up to appropriate integrations by parts.\\

It is worth noting that up to integrations by parts (i.e. up to the Gauss--Bonnet term which is irrelevant in four--dimensions), the $R^2$ terms can be expressed in terms of the Weyl term $W_{abcd}$ as follows,
\ba
\label{eq:R21loopScalar}
\frac{1}{240}\(\bar I-I_i\)\sqrt{-g}\(\frac 12 R^2+R\mn^2\)=\frac{1}{240}\(\bar I-I_i\)\sqrt{-g}\(\frac 58 R^2+\frac 14 W^2\)\,,
\ea
corresponding to the following expression
\ba
\label{eq:R2scalar}
\frac{8}{5} \Delta C_{R^2}= 4 \Delta C_{W^2} = \frac{1}{240 }\(\frac{1}{8\pi^2 \epsilon}-\frac{1}{16\pi^2}\(-1+\gamma+\log\(\frac{M^2}{4\pi \mu^2}\)\)\)\,.
\ea
Had this been calculated directly in a cutoff scheme, with cutoff $\Lambda$, we would have found
\ba
\label{eq:R2scalar}
\frac{8}{5} \Delta C_{R^2}= 4 \Delta C_{W^2} \sim  \frac{1}{240 }\(\frac{1}{16\pi^2}\(\log\(\frac{\Lambda^2}{M^2}\)\)\)\,.
\ea
which given the requirement $M \ll \Lambda$ implies $\Delta C_{W^2}>0$.

\section{Tensor Modes on FLRW from Dimension--6 Operators}
\label{app2:FLRW}

In this appendix, we shall derive the equation of motion for tensor modes coming from dimension--6 $R^3$ corrections arising from integrating out loops of massive fields of various spins. Unlike the dimension--4 operators explored previously, these are finite and independent of any renormalization procedure. Once again we work in conformal time $\eta$, with metric $\gamma\mn=a^2 \eta\mn$  and introduce the transverse and traceless tensor fluctuations $h_{ij}=\sum_\sigma h_\sigma \varepsilon^\sigma_{ij}$. Without loss of generality, we omit any mention of those two polarizations and simply denote the tensor modes as $h$ and use the normalization $g_{ij}=\gamma_{ij}+a h_{ij}$.   \\

As explained previously, the standard contributions from GR (the Einstein--Hilbert term and the low--energy matter fields) lead to the following contributions to the  tensor modes on FLRW,
\ba
\label{eq:EHm}
\L_{\rm EH, \psi}^{(hh)}=\frac{\mpl^2}{4}a^2   h  \(\frac 1{a^2}\Box_\eta +2 H^2+\dot H\) h \,,
\ea
The contributions from the dimension--6 operators are
\ba
\label{eq:Ldim6a}
\L_{\rm dim-6}^{(hh)}=\frac{1}{12(2\pi)^4}\frac {a^2}{420 M^2}\ h \ \hat{O}_{\rm dim-6} \ h\,,
\ea
where $\hat{O}_{\rm dim-{6}}$ is a  4th order operators similar in spirit to the one found in \eqref{eq:Odim41a} but with slightly different coefficients
\ba
\label{eq:Odim61a}
\hat{O}_{\rm dim-6}=\sum_{s=0,1/2,1} N^*_{s} \Bigg[
\frac 1{a^4}f_{1}^{(s)} \Box^2_\eta
+\frac 1{a^4}f_{2}^{(s)} \Box_\eta  \vec{\nabla}^2
+\frac 1{a^3}f_{3}^{(s)} \Box_\eta  \p_{\eta}
+\frac 1{a^2}f_{4}^{(s)} \Box_\eta\\
+\frac 1{a} f_{5}^{(s)} \p_{\eta}
+\frac 1{a^3} f_{6}^{(s)} \vec{\nabla}^2 \p_{\eta}
+\frac 1{a^2} f_{7}^{(s)} \vec{\nabla}^2
+ f_{8}^{(s)}
\Bigg]\nn
\ea
where the functions $f_n^{(s)}$ are function of time depend on the spin of the particle integrated out.

For scalars,
\ba
\label{eq:f10}
f_{1}^{(0)}&=& 49 H^2 + 37 \dot H{}\\
f_{2}^{(0)}&=& -18  \dot H{}\\
f_{3}^{(0)}&=& 196 H^3 - 48 H  \dot H{} - 74 \ddot H\\
f_{4}^{(0)}&=& 394 H^4 + 1128 H^2 \dot H{} + 59 H  \ddot H{} -
    37  \dddot H{} + 199 (\dot H)^2\\
f_{5}^{(0)}&=& -1968 H^3 \dot H{} - 810 H^2 \ddot H{} -
     46 H   \dddot H{} - 1620 H  (\dot H)^2 - 492 \dot H{} \ddot H{}\\
f_{6}^{(0)}&=&18 (\ddot H{} - 2 \dot H{} H )\\
f_{7}^{(0)}&=& 10 (\dot H)^2 + 46 \dddot H{} + 66 \ddot H{} H  -
     344 \dot H{} H^2\\
f_{8}^{(0)}&=& 1528 H^6 - 1608 \dot H{}^3 - 466 (\ddot H{})^2 - 466 \dot H{} \dddot H{} \\
&-& 5830 \dot H{} \ddot H{} H
- 11313 (\dot H)^2 H^2 - 841 \dddot H{} H^2 \nn
- 4443 \ddot H{} H^3 - 2836 \dot H{} H^4\,.
\ea
For vectors,
\ba
f_{1}^{(1)}&=& - 161 H^2+363 \dot H{} \\
f_{2}^{(1)}&=& -446 \dot H{}\\
f_{3}^{(1)}&=& - 644 H^3-726 \ddot H{} + 2096 \dot H{} H \\
f_{4}^{(1)}&=& - 358 H^4+1059 (\dot H)^2 - 363 \dddot H{} + 1577 \ddot H{} H -
     2300 \dot H{} H^2 \\
f_{5}^{(1)}&=& 572 (\dot H)^2 H - 166 \dddot H{} H +
     286 \ddot H{} H^2 - 4 \dot H{} (47 \ddot H{} - 680 H^3)\\
f_{6}^{(1)}&=& 446 \ddot H{} - 892 \dot H{} H\\
f_{7}^{(1)}&=& -362 (\dot H)^2 + 166 \dddot H{} - 1286 \ddot H{} H +
     872 \dot H{} H^2\\
f_{8}^{(1)}&=& - 1016 H^6 + 618 (\ddot H)^2 + 11253 (\dot H)^2 H^2 +
     109 \dddot H{} H^2 \\
     &+& 2687 \ddot H{} H^3 +1714 (\dot H)^3 +
     2 \dot H{} (309 \dddot H{} + 2441 \ddot H{} H + 1528 H^4)\nn
\ea
and finally for spinors
\ba
f_{1}^{(1/2)}&=& \frac 12  (- 21 H^2+104 \dot H{} )\\
f_{2}^{(1/2)}&=& -76 \dot H{}\\
f_{3}^{(1/2)}&=& - 42 H^3 -104 \ddot H{} + 250 \dot H{} H \\
f_{4}^{(1/2)}&=& \frac12 (- 113 H^4+359 (\dot H)^2 - 104 \dddot H{} + 450 \ddot H{} H -
     284 \dot H{} H^2 )\\
f_{5}^{(1/2)}&=& -92 (\dot H)^2 H - 48 \dddot H{} H - 46 \ddot H{} H^2 -
     \dot H{} (157 \ddot H{} - 310 H^3)\\
f_{6}^{(1/2)}&=& 76 \ddot H{} - 152 \dot H{} H\\
f_{7}^{(1/2)}&=& -90 (\dot H)^2 + 48 \dddot H{} - 160 \ddot H{} H +
     2 \dot H{} H^2\\
     \label{eq:f8sp}
f_{8}^{(1/2)}&=& \frac 12 \Big(-288 H^6 +100 (\ddot H)^2 + 2811 (\dot H)^2 H^2 +
     53 \dddot H{} H^2 + 755 \ddot H{} H^3 \\
     &+& 391 (\dot H)^3 +
     \dot H{} (100 \dddot H{} + 1088 \ddot H{} H + 753 H^4)\Big)\,.\nn
\ea

Working perturbatively in the dimension--6 operators, we may substitute the relation for $\Box_\eta h$ in terms of $h$ as derived from \eqref{eq:EHm},
\ba
\Box_\eta h = a^2(-2H^2-\dot H)h\,.
\ea
This perturbative substitution can be performed on the first line of the Operator $\hat{O}_{\rm dim-6}$ defined in \eqref{eq:Odim61a} so that only the second line remains which slightly altered coefficients,
\ba
\label{eq:Odim62a}
\hat{O}_{\rm dim-6}=\sum_{s=0,1/2,1} N^*_{s} \Bigg[
\frac 1{a} \tilde f_{5}^{(s)} \p_{\eta}
+\frac 1{a^3} \tilde f_{6}^{(s)} \vec{\nabla}^2 \p_{\eta}
+\frac 1{a^2} \tilde f_{7}^{(s)} \vec{\nabla}^2
+ \tilde f_{8}^{(s)}
\Bigg]\,,
\ea
with
\ba
\label{eq:ft5}
\tilde f_{5}^{(s)}&=&f_5^{(s)}+2 f_1^{(s)}(4H^3+6H \dot H+\ddot H)+f_3^{(s)}(-2H^2-\dot H)\\
\tilde f_{6}^{(s)}&=&f_6^{(s)}\\
\tilde f_{7}^{(s)}&=&f_7^{(s)}+f_2 (-2H^2-\dot H)\\
\tilde f_{8}^{(s)}&=&f_8^{(s)}+f_1^{(s)} (16H^4+34H^2\dot H+9H \ddot H+\dddot H+7(\dot H)^2)\\
&+&f_3^{(s)}(-4H^3-6 H \dot H-\ddot H)
+f_4^{(s)}(-2H^2-\dot H)\nn\,.
\label{eq:ft8}
\ea

The original normalization of the tensor modes $g_{ij}=a h_{ij}$ was chosen precisely so as to remove the friction term that would otherwise have arose from the standard Einstein-Hilbert term. We now perform a subleading rescaling of the tensor modes so as to absorb the subleading friction term $(\tilde f_{5}^{(s)} + a^{-2} \tilde f_{6}^{(s)} \vec{\nabla}^2)$ present in \eqref{eq:Odim62a}. For this it is easier to move to momentum space and perform the rescaling
\ba
\label{eq:FR1}
h_k(\eta)\to \(1+ \frac{\Omega(k,\eta)}{ 12\times 420 (2\pi)^4  M^2 \mpl^2} \)h_k(\eta)\,,
\ea
with
\ba
\label{eq:FR2}
\Omega(k,\eta)&=&-4\(9\Ns+38\Np+223 \Nv \)\dot H\frac{k^2}{a^2}
- (592 \Ns - 71 \Np - 72 \Nv)H^4\\
&-& 14 (74 \Ns -27 \Np -226 \Nv) H^2 \dot H
-(298 \Ns -99 \Np + 1430 \Nv) (\dot H)^2 \nn \\
&-&
(92 \Ns + 96 \Np + 332 \Nv) H \ddot H  \,,\nn
\ea
leading to the following equation of motion for the tensor modes,
\ba
h'' +c_s^2 k^2 h= m_0^2 h\,,
\ea
with
\ba
\label{eq:cs}
c_s^2=1-\frac{1}{12\times 105 (2\pi)^4  M^2 \mpl^2}&\Big(&
 2 (163 \Ns - 39 \Np - 659 \Nv)  H^2 \dot H\\
&-& (46 \Ns + 62 \Np + 530 \Nv) \dot H^2\nn\\
&+& (-93 \Ns + 46 \Np + 617 \Nv) H \ddot H\nn\\
&+& (-37 \Ns - 10 \Np + 57 \Nv) \dddot H
\Big)\nn\,,
\ea
while the IR  part in the dispersion relation is
\ba
m_0^2=a^2 \( -2H^2-\dot H+\mathcal{O}\(\frac{H^6}{\mpl^2 M^2}\) \)\,.
\ea

\section{Resumming Green's Function Secular Behaviour}

\label{app:resumming}

Rather than following the logic presented in Section~\ref{sec:speedID} to determine the speed, one could have attempted to define the Green's function iteratively around the GR one. Following such an approach, one would be lured into the belief that the causal structure appears to be the standard one at any finite order.  A similar argument can be applied to any theory with a small departure of the sound speed. The reason why such a procedure is not correct in general is because it relies on perturbations that carry secular effects, whose growth can in turn undermine the perturbative expansion. \\

Indeed in rewriting in the form \eqref{eq3} we have implicitly resummed effects, similar to a self energy resummation, which contribute already at two--derivative order. Superficially higher order time derivatives generate second order time derivatives by partly acting on the background,
e.g.
\be
\partial^n ( a(\eta)^m h ) = \dots \frac{n!}{2! (n-2)!} (\partial^{n-2} a(\eta)^m)  \partial^2 h  + \dots \, ,
\ee
hence giving contribution to the low energy sound speed. The superficially larger contributions from higher time derivatives are cancelled by similar spatial derivative terms by virtue of the leading order equations of motion.
Field redefining the equation of motion into standard two time derivative form \eqref{eq3} is helpful so that we may define the retarded propagator in the standard way via time-ordering, i.e.
$G_{\rm ret}(x,x') =i \theta(t-t') \Delta(x,x') = i \langle   \hat T h(x) h(x') \rangle-i \langle  h(x) h(x') \rangle$ with $ \Delta(x,x')$ the commutator function. But perhaps more importantly the resummation removes secular behaviours that would have been obtained otherwise. Specifically the resummed WKB modes can be rewritten perturbatively in terms of the unresummed ones,
\be
e^{i k x- i \int^{\eta} \d \eta' k/c_s(\eta)}  = e^{i k x-i\int^{\eta} \d \eta' k} ( 1+ i k \int^{\eta} \d \eta' (c_s(\eta')-1) + \dots)\,.
\ee
The secular perturbative growth of the right hand side would become large for $|k \eta | \sim 1/(c_s-1)$, invalidating this perturbative approach  of solving \eqref{eq2} by directly iterating about the GR result. Instead, in Section~\ref{sec:speedID}, we have identified the retarded Green's function by iterating the resummed counterpart given in Eqns.~(\ref{eq:iterating2}/\ref{eq:iterating1}). In doing so, we have effectively resummed an infinite number of contributions, leading to a better behaved perturbative expansion. This is the approach we follow in identifying the speed throughout this work. \\

Ultimately the causalstructure is determined by the front velocity, i.e. the high energy limit of the phase velocity. As we have argued we expect this to be luminal in a fundamentally Lorentz invariant theory.

\bibliographystyle{JHEP}
\bibliography{references}

\end{document}